\documentclass[lettersize,journal]{IEEEtran}

\usepackage[utf8]{inputenc}
\usepackage{cite}
\usepackage{soul, color}
\usepackage{graphicx, lipsum}
\usepackage{subcaption}
\usepackage{float}
\usepackage[colorlinks=true, urlcolor=blue, linkcolor=red]{hyperref}
\usepackage{booktabs}
\usepackage{adjustbox}
\usepackage{amsmath}
\usepackage{float}
\usepackage{graphicx}
\usepackage{array}
\usepackage{fancyhdr}
\usepackage{filemod}

\usepackage{rotating}

\usepackage{stfloats}

\usepackage[dvipsnames]{xcolor}

\usepackage[acronyms,nonumberlist,nopostdot,nomain,nogroupskip,acronymlists={hidden}]{glossaries}
\newglossary[algh]{hidden}{acrh}{acnh}{Hidden Acronyms}
\glsdisablehyper
\newacronym{3gpp}{3GPP}{3rd Generation Partnership Project}
\newacronym{4g}{4G}{4th generation mobile network}
\newacronym{5g}{5G}{5th generation mobile network}
\newacronym{6g}{6G}{6th generation mobile network}
\newacronym{nextg}{NextG}{Next Generation}
\newacronym{5gc}{5GC}{5G Core}
\newacronym{adc}{ADC}{Analog to Digital Converter}
\newacronym{aerpaw}{AERPAW}{Aerial Experimentation and Research Platform for Advanced Wireless}
\newacronym{ai}{AI}{Artificial Intelligence}
\newacronym{aimd}{AIMD}{Additive Increase Multiplicative Decrease}
\newacronym{am}{AM}{Acknowledged Mode}
\newacronym{amc}{AMC}{Adaptive Modulation and Coding}
\newacronym{amf}{AMF}{Access and Mobility Management Function}
\newacronym{aops}{AOPS}{Adaptive Order Prediction Scheduling}
\newacronym{api}{API}{Application Programming Interface}
\newacronym{apn}{APN}{Access Point Name}
\newacronym{aqm}{AQM}{Active Queue Management}
\newacronym{ausf}{AUSF}{Authentication Server Function}
\newacronym{avc}{AVC}{Advanced Video Coding}
\newacronym{awgn}{AGWN}{Additive White Gaussian Noise}
\newacronym{balia}{BALIA}{Balanced Link Adaptation Algorithm}
\newacronym{bbu}{BBU}{Base Band Unit}
\newacronym{bdp}{BDP}{Bandwidth-Delay Product}
\newacronym{ber}{BER}{Bit Error Rate}
\newacronym{bf}{BF}{Beamforming}
\newacronym{bler}{BLER}{Block Error Rate}
\newacronym{brr}{BRR}{Bayesian Ridge Regressor}
\newacronym{bsr}{BSR}{Buffer Status Report}
\newacronym{bs}{BS}{Base Station}
\newacronym{bpsk}{BPSK}{Binary Phase-shift keying}
\newacronym{bss}{BSS}{Business Support System}
\newacronym{ca}{CA}{Carrier Aggregation}
\newacronym{caas}{CaaS}{Connectivity-as-a-Service}
\newacronym{cb}{CB}{Code Block}
\newacronym{cc}{CC}{Congestion Control}
\newacronym{ccid}{CCID}{Congestion Control ID}
\newacronym{cco}{CC}{Carrier Component}
\newacronym{cd}{CD}{Continuous Delivery}
\newacronym{cdd}{CDD}{Cyclic Delay Diversity}
\newacronym{cdf}{CDF}{Cumulative Distribution Function}
\newacronym{cdma}{CDMA}{Code-Division Multiple Access}
\newacronym{cdn}{CDN}{Content Distribution Network}
\newacronym{ci}{CI}{Continuous Integration}
\newacronym{cicd}{CI/CD}{Continuous Integration/Continuous Delivery}
\newacronym{cir}{CIR}{Channel Impulse Response}
\newacronym{cn}{CN}{Core Network}
\newacronym{codel}{CoDel}{Controlled Delay Management}
\newacronym{comac}{COMAC}{Converged Multi-Access and Core}
\newacronym{cord}{CORD}{Central Office Re-architected as a Datacenter}
\newacronym{cornet}{CORNET}{COgnitive Radio NETwork}
\newacronym{cosmos}{COSMOS}{Cloud Enhanced Open Software Defined Mobile Wireless Testbed for City-Scale Deployment}
\newacronym{cots}{COTS}{Commercial Off-the-Shelf}
\newacronym{cp}{CP}{Control Plane}
\newacronym{cpu}{CPU}{Central Processing Unit}
\newacronym{cqi}{CQI}{Channel Quality Information}
\newacronym{cr}{CR}{Cognitive Radio}
\newacronym{cran}{CRAN}{Cloud \gls{ran}}
\newacronym{crs}{CRS}{Cell Reference Signal}
\newacronym{csi}{CSI}{Channel State Information}
\newacronym{csirs}{CSI-RS}{Channel State Information - Reference Signal}
\newacronym{cu}{CU}{Central Unit}
\newacronym{d2tcp}{D$^2$TCP}{Deadline-aware Data center TCP}
\newacronym{d3}{D$^3$}{Deadline-Driven Delivery}
\newacronym{dac}{DAC}{Digital to Analog Converter}
\newacronym{dag}{DAG}{Directed Acyclic Graph}
\newacronym{darpa}{DARPA}{Defense Advanced Research Projects Agency}
\newacronym{das}{DAS}{Distributed Antenna System}
\newacronym{dash}{DASH}{Dynamic Adaptive Streaming over HTTP}
\newacronym{dbs}{DBS}{Deep Brain Stimulation}
\newacronym{dc}{DC}{Dual Connectivity}
\newacronym{dccp}{DCCP}{Datagram Congestion Control Protocol}
\newacronym{dce}{DCE}{Direct Code Execution}
\newacronym{dci}{DCI}{Downlink Control Information}
\newacronym{dcl}{DCL}{Dear Colleague Letter}
\newacronym{dctcp}{DCTCP}{Data Center TCP}
\newacronym{devops}{DevOps}{Development and Operations}
\newacronym{dl}{DL}{Deep Learning}
\newacronym{dmr}{DMR}{Deadline Miss Ratio}
\newacronym{dmrs}{DMRS}{DeModulation Reference Signal}
\newacronym{drlcc}{DRL-CC}{Deep Reinforcement Learning Congestion Control}
\newacronym{drs}{DRS}{Discovery Reference Signal}
\newacronym{dt}{DT}{Digital Twin}
\newacronym{dtn}{DTN}{Digital Twin Network}
\newacronym{dtmn}{DTMN}{Digital Twin for Mobile Network}
\newacronym{dtwn}{DTWN}{Digital Twin Wireless Network}
\newacronym{du}{DU}{Distributed Unit}
\newacronym{e2e}{E2E}{end-to-end}
\newacronym{ecaas}{ECaaS}{Edge-Cloud-as-a-Service}
\newacronym{ecn}{ECN}{Explicit Congestion Notification}
\newacronym{edf}{EDF}{Earliest Deadline First}
\newacronym{em}{EM}{Electro-Magnetic}
\newacronym{embb}{eMBB}{Enhanced Mobile Broadband}
\newacronym{empower}{EMPOWER}{EMpowering transatlantic PlatfOrms for advanced WirEless Research}
\newacronym{enb}{eNB}{evolved Node Base}
\newacronym{endc}{EN-DC}{E-UTRAN-\gls{nr} \gls{dc}}
\newacronym{epc}{EPC}{Evolved Packet Core}
\newacronym{eps}{EPS}{Evolved Packet System}
\newacronym{es}{ES}{Edge Server}
\newacronym{etsi}{ETSI}{European Telecommunications Standards Institute}
\newacronym[firstplural=Estimated Times of Arrival (ETAs)]{eta}{ETA}{Estimated Time of Arrival}
\newacronym{eutran}{E-UTRAN}{Evolved Universal Terrestrial Access Network}
\newacronym{faas}{FaaS}{Function-as-a-Service}
\newacronym{fapi}{FAPI}{Functional Application Platform Interface}
\newacronym{fcc}{FCC}{Federal Communications Commission}
\newacronym{fdd}{FDD}{Frequency Division Duplexing}
\newacronym{fdm}{FDM}{Frequency Division Multiplexing}
\newacronym{fdma}{FDMA}{Frequency Division Multiple Access}
\newacronym{fed4fire}{FED4FIRE+}{Federation 4 Future Internet Research and Experimentation Plus}
\newacronym{fir}{FIR}{Finite Impulse Response}
\newacronym{fit}{FIT}{Future \acrlong{iot}}
\newacronym{fl}{FL}{Federated Learning}
\newacronym{fpga}{FPGA}{Field Programmable Gate Array}
\newacronym{fr2}{FR2}{Frequency Range 2}
\newacronym{fs}{FS}{Fast Switching}
\newacronym{fscc}{FSCC}{Flow Sharing Congestion Control}
\newacronym{ftp}{FTP}{File Transfer Protocol}
\newacronym{fw}{FW}{Flow Window}
\newacronym{ga128}{Ga}{Golay Sequence type A}
\newacronym{ge}{GE}{Gaussian Elimination}
\newacronym{glfsr}{GLFSR}{Galois Linear Feedback Shift Register}
\newacronym{gnb}{gNB}{Next Generation Node Base}
\newacronym{gold}{Gold}{Gold}
\newacronym{gop}{GOP}{Group of Pictures}
\newacronym{gpr}{GPR}{Gaussian Process Regressor}
\newacronym{gpu}{GPU}{Graphics Processing Unit}
\newacronym{gtp}{GTP}{GPRS Tunneling Protocol}
\newacronym{gtpc}{GTP-C}{GPRS Tunnelling Protocol Control Plane}
\newacronym{gtpu}{GTP-U}{GPRS Tunnelling Protocol User Plane}
\newacronym{gtpv2c}{GTPv2-C}{\gls{gtp} v2 - Control}
\newacronym{gw}{GW}{Gateway}
\newacronym{harq}{HARQ}{Hybrid Automatic Repeat reQuest}
\newacronym{hetnet}{HetNet}{Heterogeneous Network}
\newacronym{hh}{HH}{Hard Handover}
\newacronym{hol}{HOL}{Head-of-Line}
\newacronym{hqf}{HQF}{Highest-quality-first}
\newacronym{hss}{HSS}{Home Subscription Server}
\newacronym{http}{HTTP}{HyperText Transfer Protocol}
\newacronym{ia}{IA}{Initial Access}
\newacronym{iab}{IAB}{Integrated Access and Backhaul}
\newacronym{ic}{IC}{Incident Command}
\newacronym{ietf}{IETF}{Internet Engineering Task Force}
\newacronym{ifw}{IFW}{Interference Free Window}
\newacronym{imsi}{IMSI}{International Mobile Subscriber Identity}
\newacronym{imt}{IMT}{International Mobile Telecommunication}
\newacronym{iot}{IoT}{Internet of Things}
\newacronym{ip}{IP}{Internet Protocol}
\newacronym{iq}{IQ}{In-phase and Quadrature}
\newacronym{isi}{ISI}{Inter-Symbol Interference}
\newacronym{itu}{ITU}{International Telecommunication Union}
\newacronym{kpi}{KPI}{Key Performance Indicator}
\newacronym{kvm}{KVM}{Kernel-based Virtual Machine}
\newacronym{lfsr}{LFSR}{Linear Feedback Shift Register}
\newacronym{los}{LOS}{Line-of-Sight}
\newacronym{ls}{LS}{Loosely Synchronised}
\newacronym{lsm}{LSM}{Link-to-System Mapping}
\newacronym{lstm}{LSTM}{Long Short Term Memory}
\newacronym{lte}{LTE}{Long Term Evolution}
\newacronym{lxc}{LXC}{Linux Container}
\newacronym{m2m}{M2M}{Machine to Machine}
\newacronym{mac}{MAC}{Medium Access Control}
\newacronym{mai}{MAI}{Multiple Access Interference}
\newacronym{manet}{MANET}{Mobile Ad Hoc Network}
\newacronym{mano}{MANO}{Management and Orchestration}
\newacronym{mc}{MC}{Multi-Connectivity}
\newacronym{mcc}{MCC}{Mobile Cloud Computing}
\newacronym{mchem}{MCHEM}{Massive Channel Emulator}
\newacronym{mcs}{MCS}{Modulation and Coding Scheme}
\newacronym{mec}{MEC}{Multi-access Edge Computing}
\newacronym{mec2}{MEC}{Mobile Edge Cloud}
\newacronym{mec3}{MEC}{Mobile Edge Computing}
\newacronym{mfc}{MFC}{Mobile Fog Computing}
\newacronym{mi}{MI}{Mutual Information}
\newacronym{mib}{MIB}{Master Information Block}
\newacronym{miesm}{MIESM}{Mutual Information Based Effective SINR}
\newacronym{mimo}{MIMO}{Multiple Input, Multiple Output}
\newacronym{mgen}{MGEN}{Multi-Generator}
\newacronym{ml}{ML}{Machine Learning}
\newacronym{mlr}{MLR}{Maximum-local-rate}
\newacronym[plural=\gls{mme}s,firstplural=Mobility Management Entities (MMEs)]{mme}{MME}{Mobility Management Entity}
\newacronym{mmtc}{mMTC}{Massive Machine-Type Communications}
\newacronym{mmwave}{mmWave}{millimeter wave}
\newacronym{mpdccp}{MP-DCCP}{Multipath Datagram Congestion Control Protocol}
\newacronym{mptcp}{MPTCP}{Multipath TCP}
\newacronym{mr}{MR}{Maximum Rate}
\newacronym{mrdc}{MR-DC}{Multi \gls{rat} \gls{dc}}
\newacronym{mse}{MSE}{Mean Square Error}
\newacronym{mss}{MSS}{Maximum Segment Size}
\newacronym{mt}{MT}{Mobile Termination}
\newacronym{mtd}{MTD}{Machine-Type Device}
\newacronym{mtu}{MTU}{Maximum Transmission Unit}
\newacronym{mumimo}{MU-MIMO}{Multi-user \gls{mimo}}
\newacronym{mvno}{MVNO}{Mobile Virtual Network Operator}
\newacronym{nalu}{NALU}{Network Abstraction Layer Unit}
\newacronym{nas}{NAS}{Network Attached Storage}
\newacronym{nbiot}{NB-IoT}{Narrow Band IoT}
\newacronym{nfv}{NFV}{Network Function Virtualization}
\newacronym{nfvi}{NFVI}{Network Function Virtualization Infrastructure}
\newacronym{nic}{NIC}{Network Interface Card}
\newacronym{nlos}{NLOS}{Non-Line-of-Sight}
\newacronym{now}{NOW}{Non Overlapping Window}
\newacronym{nrdz}{NRDZ}{National Radio Dynamic Zone}
\newacronym{nsf}{NSF}{National Science Foundation}
\newacronym{nsm}{NSM}{Network Service Mesh}
\newacronym[type=hidden]{nr}{NR}{New Radio}
\newacronym{nrf}{NRF}{Network Repository Function}
\newacronym{nsa}{NSA}{Non Stand Alone}
\newacronym{nse}{NSE}{Network Slicing Engine}
\newacronym{nssf}{NSSF}{Network Slice Selection Function}
\newacronym{ntp}{NTP}{Network Time Protocol}
\newacronym{o2i}{O2I}{Outdoor to Indoor}
\newacronym{oai}{OAI}{OpenAirInterface}
\newacronym{oaicn}{OAI-CN}{\gls{oai} \acrlong{cn}}
\newacronym{oairan}{OAI-RAN}{\acrlong{oai} \acrlong{ran}}
\newacronym{oam}{OAM}{Operations, Administration and Maintenance}
\newacronym[plural=\gls{obu}s,firstplural=Onboard Units (OBUs)]{obu}{OBU}{Onboard Unit}
\newacronym{ofdm}{OFDM}{Orthogonal Frequency Division Multiplexing}
\newacronym{olia}{OLIA}{Opportunistic Linked Increase Algorithm}
\newacronym{omec}{OMEC}{Open Mobile Evolved Core}
\newacronym{onap}{ONAP}{Open Network Automation Platform}
\newacronym{onf}{ONF}{Open Networking Foundation}
\newacronym{onos}{ONOS}{Open Networking Operating System}
\newacronym{oom}{OOM}{\gls{onap} Operations Manager}
\newacronym{opnfv}{OPNFV}{Open Platform for \gls{nfv}}
\newacronym[type=hidden]{oran}{O-RAN}{Open RAN}
\newacronym{orbit}{ORBIT}{Open-Access Research Testbed for Next-Generation Wireless Networks}
\newacronym{os}{OS}{Operating System}
\newacronym{osm}{OSM}{Open Street Map}
\newacronym{oss}{OSS}{Operations Support System}
\newacronym{pa}{PA}{Position-aware}
\newacronym{pase}{PASE}{Prioritization, Arbitration, and Self-adjusting Endpoints}
\newacronym{pawr}{PAWR}{Platforms for Advanced Wireless Research}
\newacronym{pbch}{PBCH}{Physical Broadcast Channel}
\newacronym{pcef}{PCEF}{Policy and Charging Enforcement Function}
\newacronym{pcfich}{PCFICH}{Physical Control Format Indicator Channel}
\newacronym{pcrf}{PCRF}{Policy and Charging Rules Function}
\newacronym{pdcch}{PDCCH}{Physical Downlink Control Channel}
\newacronym{pdcp}{PDCP}{Packet Data Convergence Protocol}
\newacronym{pdsch}{PDSCH}{Physical Downlink Shared Channel}
\newacronym{pdu}{PDU}{Packet Data Unit}
\newacronym{pdp}{PDP}{Power Delay Profile}
\newacronym{pf}{PF}{Proportional Fair}
\newacronym{pgw}{PGW}{Packet Gateway}
\newacronym{phich}{PHICH}{Physical Hybrid ARQ Indicator Channel}
\newacronym{phy}{PHY}{Physical}
\newacronym{pl}{PL}{Path Loss}
\newacronym{pmch}{PMCH}{Physical Multicast Channel}
\newacronym{pmi}{PMI}{Precoding Matrix Indicators}
\newacronym{powder}{POWDER}{Platform for Open Wireless Data-driven Experimental Research}
\newacronym{ppo}{PPO}{Proximal Policy Optimization}
\newacronym{ppp}{PPP}{Poisson Point Process}
\newacronym{prach}{PRACH}{Physical Random Access Channel}
\newacronym{prb}{PRB}{Physical Resource Block}
\newacronym{psnr}{PSNR}{Peak Signal to Noise Ratio}
\newacronym{pss}{PSS}{Primary Synchronization Signal}
\newacronym{pucch}{PUCCH}{Physical Uplink Control Channel}
\newacronym{pusch}{PUSCH}{Physical Uplink Shared Channel}
\newacronym{qam}{QAM}{Quadrature Amplitude Modulation}
\newacronym{qci}{QCI}{\gls{qos} Class Identifier}
\newacronym{qoe}{QoE}{Quality of Experience}
\newacronym{qos}{QoS}{Quality of Service}
\newacronym{qtgui}{QT-GUI}{QT Graphical User Interface}
\newacronym{quic}{QUIC}{Quick UDP Internet Connections}
\newacronym{rach}{RACH}{Random Access Channel}
\newacronym{ran}{RAN}{Radio Access Network}
\newacronym[firstplural=Radio Access Technologies (RATs)]{rat}{RAT}{Radio Access Technology}
\newacronym{rcn}{RCN}{Research Coordination Network}
\newacronym{rec}{REC}{Radio Edge Cloud}
\newacronym{red}{RED}{Random Early Detection}
\newacronym{renew}{RENEW}{Reconfigurable Eco-system for Next-generation End-to-end Wireless}
\newacronym{rf}{RF}{Radio Frequency}
\newacronym{rfc}{RFC}{Request for Comments}
\newacronym{rfr}{RFR}{Random Forest Regressor}
\newacronym{ric}{RIC}{RAN Intelligent Controller}
\newacronym{rlc}{RLC}{Radio Link Control}
\newacronym{rlf}{RLF}{Radio Link Failure}
\newacronym{rlnc}{RLNC}{Random Linear Network Coding}
\newacronym{rmse}{RMSE}{Root Mean Squared Error}
\newacronym{rnis}{RNIS}{Radio Network Information Service}
\newacronym{rr}{RR}{Round Robin}
\newacronym{rrc}{RRC}{Radio Resource Control}
\newacronym{rrm}{RRM}{Radio Resource Management}
\newacronym{rru}{RRU}{Remote Radio Unit}
\newacronym{rs}{RS}{Remote Server}
\newacronym{rsrp}{RSRP}{Reference Signal Received Power}
\newacronym{rsrq}{RSRQ}{Reference Signal Received Quality}
\newacronym{rss}{RSS}{Received Signal Strength}
\newacronym{rssi}{RSSI}{Received Signal Strength Indicator}
\newacronym{rsu}{RSU}{Road-Side Unit}
\newacronym{rtt}{RTT}{Round Trip Time}
\newacronym{ru}{RU}{Radio Unit}
\newacronym{rw}{RW}{Receive Window}
\newacronym{rx}{RX}{Receiver}
\newacronym{s1ap}{S1AP}{S1 Application Protocol}
\newacronym{sa}{SA}{standalone}
\newacronym{sack}{SACK}{Selective Acknowledgment}
\newacronym{sap}{SAP}{Service Access Point}
\newacronym{sc2}{SC2}{Spectrum Collaboration Challenge}
\newacronym{scef}{SCEF}{Service Capability Exposure Function}
\newacronym{sch}{SCH}{Secondary Cell Handover}
\newacronym{scoot}{SCOOT}{Split Cycle Offset Optimization Technique}
\newacronym{sctp}{SCTP}{Stream Control Transmission Protocol}
\newacronym{sdap}{SDAP}{Service Data Adaptation Protocol}
\newacronym{sd}{SD}{Standard Deviation}
\newacronym{sdk}{SDK}{Software Development Kit}
\newacronym{sdm}{SDM}{Space Division Multiplexing}
\newacronym{sdma}{SDMA}{Spatial Division Multiple Access}
\newacronym{sdn}{SDN}{Software-defined Networking}
\newacronym{sdr}{SDR}{Software-defined Radio}
\newacronym{seba}{SEBA}{SDN-Enabled Broadband Access}
\newacronym{sgsn}{SGSN}{Serving GPRS Support Node}
\newacronym{sgw}{SGW}{Service Gateway}
\newacronym{si}{SI}{Study Item}
\newacronym{sib}{SIB}{Secondary Information Block}
\newacronym{sinr}{SINR}{Signal to Interference plus Noise Ratio}
\newacronym{sip}{SIP}{Session Initiation Protocol}
\newacronym{siso}{SISO}{Single Input, Single Output}
\newacronym{sla}{SLA}{Service Level Agreement}
\newacronym{sm}{SM}{Saturation Mode}
\newacronym{smf}{SMF}{Session Management Function}
\newacronym{smo}{SMO}{Service Management and Orchestration}
\newacronym{sms}{SMS}{Short Message Service}
\newacronym{smsgmsc}{SMS-GMSC}{\gls{sms}-Gateway}
\newacronym{snr}{SNR}{Signal-to-Noise-Ratio}
\newacronym{son}{SON}{Self-Organizing Network}
\newacronym{sptcp}{SPTCP}{Single Path TCP}
\newacronym{srb}{SRB}{Service Radio Bearer}
\newacronym{srn}{SRN}{Standard Radio Node}
\newacronym{srs}{SRS}{Sounding Reference Signal}
\newacronym{ss}{SS}{Synchronization Signal}
\newacronym{sss}{SSS}{Secondary Synchronization Signal}
\newacronym{st}{ST}{Spanning Tree}
\newacronym{svc}{SVC}{Scalable Video Coding}
\newacronym{tb}{TB}{Transport Block}
\newacronym{tcp}{TCP}{Transmission Control Protocol}
\newacronym{tdd}{TDD}{Time Division Duplexing}
\newacronym{tdm}{TDM}{Time Division Multiplexing}
\newacronym{tdma}{TDMA}{Time Division Multiple Access}
\newacronym{tfl}{TfL}{Transport for London}
\newacronym{tfrc}{TFRC}{TCP-Friendly Rate Control}
\newacronym{tft}{TFT}{Traffic Flow Template}
\newacronym{tgen}{TGEN}{Traffic Generator}
\newacronym{tip}{TIP}{Telecom Infra Project}
\newacronym{tm}{TM}{Transparent Mode}
\newacronym{to}{TO}{Telco Operator}
\newacronym{toa}{ToA}{Time of Arrival}
\newacronym{tr}{TR}{Technical Report}
\newacronym{trp}{TRP}{Transmitter Receiver Pair}
\newacronym{ts}{TS}{Technical Specification}
\newacronym{tti}{TTI}{Transmission Time Interval}
\newacronym{ttt}{TTT}{Time-to-Trigger}
\newacronym{tx}{TX}{Transmitter}
\newacronym{uas}{UAS}{Unmanned Aerial System}
\newacronym{uav}{UAV}{Unmanned Aerial Vehicle}
\newacronym{udm}{UDM}{Unified Data Management}
\newacronym{udp}{UDP}{User Datagram Protocol}
\newacronym{udr}{UDR}{Unified Data Repository}
\newacronym{ue}{UE}{User Equipment}
\newacronym{uhd}{UHD}{\gls{usrp} Hardware Driver}
\newacronym{ul}{UL}{Uplink}
\newacronym{um}{UM}{Unacknowledged Mode}
\newacronym{uml}{UML}{Unified Modeling Language}
\newacronym{upa}{UPA}{Uniform Planar Array}
\newacronym{upf}{UPF}{User Plane Function}
\newacronym{urllc}{URLLC}{Ultra Reliable and Low Latency Communication}
\newacronym{usa}{U.S.}{United States}
\newacronym{usim}{USIM}{Universal Subscriber Identity Module}
\newacronym{usrp}{USRP}{Universal Software Radio Peripheral}
\newacronym{utc}{UTC}{Urban Traffic Control}
\newacronym{vim}{VIM}{Virtualization Infrastructure Manager}
\newacronym{vm}{VM}{Virtual Machine}
\newacronym{vnf}{VNF}{Virtual Network Function}
\newacronym{volte}{VoLTE}{Voice over \gls{lte}}
\newacronym{voltha}{VOLTHA}{Virtual OLT HArdware Abstraction}
\newacronym{vr}{VR}{Virtual Reality}
\newacronym{vran}{vRAN}{Virtualized \gls{ran}}
\newacronym{vss}{VSS}{Video Streaming Server}
\newacronym{wbf}{WBF}{Wired Bias Function}
\newacronym{wf}{WF}{Wired-first}
\newacronym{wi}{WI}{Wireless InSite}
\newacronym{wlan}{WLAN}{Wireless Local Area Network}
\newacronym{pnf}{PNF}{Physical Network Function}
\newacronym{drl}{DRL}{Deep Reinforcement Learning}
\newacronym{mtc}{MTC}{Machine-type Communications}
\newacronym{v2x}{V2X}{Vehicle-to-everything}
\newacronym{cast}{CaST}{Channel emulation scenario generator and Sounder Toolchain}
\newacronym{gui}{GUI}{Graphical User Interface}
\newacronym{ups}{UPS}{Uninterruptible Power Supply}
\newacronym{ota}{OTA}{Over-the-Air}
\newacronym{hitl}{HITL}{hardware-in-the-loop}
\newacronym{soc}{SoC}{System-on-Chip}
\newacronym{eeg}{EEG}{electroencephalogram}
\newacronym{ieeg}{iEEG}{intracranial electroencephalogram}
\newacronym{ecg}{ECG}{electrocardiogram}
\newacronym{fph}{FPH}{false positive per hour}
\newacronym{cnn}{CNN}{Convolutional Neural Network}
\newacronym{ban}{BAN}{Body Area Network}
\newacronym{roc}{ROC}{Receiver Operating Characteristic Curve}
\newacronym{auc}{AUC}{Area Under the Curve}
\newacronym{ecdf}{eCDF}{Empirical Cumulative Distribution Function}
\newacronym{deepchem}{DeepChEmM}{Deep Learning Channel Emulation Model}
\newacronym{svm}{SVM}{Support Vector Machine}
\newacronym{rbf}{RBF}{Radial Basis Function}
\newacronym{knn}{KNN}{$k$-Nearest Neighbors}
\newacronym{gbsm}{GBSM}{Geometry-Based Stochastic Model}
\newacronym{mlp}{MLP}{Multilayer Perceptron}
\newacronym{fc}{FC}{Fully Connected}
\newacronym{ae}{AE}{Autoencoder}
\newacronym{olla}{OLLA}{Outer Loop Link Adaptation}

\iftrue
\usepackage{tikzpagenodes,etoolbox}
\usetikzlibrary{calc}
\usepackage[contents={}]{background}
\AddEverypageHook{%
\ifnumequal{\thepage}{1}{%
    \tikz[remember picture,overlay]{%
        \node[draw,
        minimum width=0.9\textwidth,
        text width=0.9\textwidth,
        font=\footnotesize
        ]
        at ($(current page header area) - (0,-1pt)$)
        {%
        This work has been accepted for publication on IEEE Transactions on Wireless Communications.\\
        ©2026 IEEE. Personal use of this material is permitted. Permission from IEEE must be obtained for all other uses, in any current or future media, including reprinting/republishing this material for advertising or promotional purposes, creating new collective works, for resale or redistribution to servers or lists, or reuse of any copyrighted component of this work in other works.
        };
    }%
}{}
}
\fi

\newcommand{\review}[1]{#1}

\begin{document}
\title{AIRMap: AI-Generated Radio Maps \\
for Wireless Digital Twins
\thanks{This work was supported in part by VIAVI Solutions, Inc.; by the Public Wireless Supply Chain Innovation Fund (PWSCIF) of the National Telecommunications and Information Administration (NTIA) under Award No. 25-60-IF011; and by the U.S. National Science Foundation under Grant CNS-1925601.}
}

\author{\IEEEauthorblockN{Ali Saeizadeh$^\dagger$, Miead Tehrani-Moayyed$^\dagger$, Davide Villa$^\dagger$, J.\ Gordon Beattie, Jr.$^*$, 
\\Pedram Johari$^\dagger$, Stefano Basagni$^\dagger$, Tommaso Melodia$^\dagger$}

\IEEEauthorblockA{$^\dagger$Institute for Intelligent Networked Systems, Northeastern University, Boston, MA, U.S.A.\\
$^*$VIAVI Solutions, Inc.\\
E-mail: $^\dagger$\{saeizadeh.a, tehranimoayyed.m, villa.d, p.johari, s.basagni, melodia\}@northeastern.edu, \\$^*$gordon.beattiejr@viavisolutions.com
}}
\maketitle

\begin{abstract}
Accurate, low-latency channel modeling is essential for real‑time wireless network simulation and digital‑twin applications. %
Traditional modeling methods like ray tracing are however computationally demanding and unsuited to model dynamic conditions. 
In this paper, we propose AIRMap, a deep‑learning framework for ultra‑fast radio‑map estimation, along with an automated pipeline for creating the largest radio‑map dataset to date. 
AIRMap uses a single‑input U‑Net autoencoder that processes only a 2D elevation map of terrain and building heights. 
\review{
Trained on 1.2M Boston-area samples and validated across four distinct urban and rural environments with varying terrain and building density, AIRMap predicts path gain with under 4\,dB RMSE in 4\,ms per inference on an NVIDIA L40S —over $100\times$ faster than GPU‑accelerated ray tracing based radio maps.
}
\review{
A lightweight calibration using just 20\% of field measurements reduces the median error to approximately 5\%, significantly outperforming traditional simulators, which exceed 50\% error.
}
Integration into the Colosseum emulator and the Sionna SYS platform demonstrate near-zero error in spectral efficiency and block‑error rate compared to measurement‑based channels. 
These findings validate AIRMap’s potential for scalable, accurate, and real‑time radio map estimation in wireless digital twins.
\end{abstract}

\section{Introduction}
\label{sec:intro}


Digital twins have emerged as transformative tools in wireless networks, creating virtual replicas—or ``multiverse''—of physical deployments that operate alongside the real world. Real‑time digital twins let developers design, test, and evaluate ``what‑if'' scenarios without risking live operations~\cite{alkhateeb_real-time_2023, noauthor_ericsson}. Furthermore, by operating ahead of real time, they can forecast potential events, and in a fully closed‑loop implementation, continuously ingest live data, update their models, and even trigger actions back in the physical network.

Real‑time digital twins support a broad spectrum of wireless network use cases by leveraging continuously updated environmental models and fast inference engines. For instance, a \gls{fdd} massive MIMO base station can offload downlink channel estimation—or at least its dominant subspace—to the digital twin, dramatically reducing pilot and feedback overhead in the physical layer~\cite{guo_deep_2024}. In the radio interface, mobile users benefit from proactive \gls{los} blockage prediction: by tracking scatterer motion in the twin, the system can initiate beam handovers or reconfiguration before link degradation occurs~\cite{alrabeiah_deep_2020}. At the network layer, gathering site-specific datasets and sensing can be used to train ML models that achieve over 90\% top‑2 beam prediction accuracy with minimal real‐world fine‑tuning, enabling rapid beam management and resource allocation~\cite{elsayed_radio_2020}. Finally, at the application layer, predicted link disruptions drive pre‑caching strategies—for example, buffering video segments ahead of an anticipated outage—to maintain seamless user experiences under stringent latency constraints~\cite{li_caas_2017}. These examples demonstrate how real‑time digital twins enable end‑to‑end optimization and adaptive control across all layers of wireless networks.

While traditional network simulators and digital twins both create virtual representations of networks, they fundamentally differ in their relationship with physical systems. Conventional simulators operate in isolation, running predefined scenarios with static inputs to predict theoretical outcomes~\cite{belldigital}. In contrast, digital twin networks establish a bidirectional connection with their physical counterparts, continuously ingesting real-time data to create high-fidelity representations that evolve alongside the actual network~\cite{robinson_twinet_2024}. This interactive mapping enables digital twins to not only reflect the current state of the network but also to provide closed-loop automation capabilities, where changes validated in the virtual environment can be safely applied to the physical network. This real-time synchronization capability makes digital twins particularly valuable for mission-critical applications in next-generation wireless networks, where they can facilitate network optimization, predictive maintenance, and innovative service development without risking operational disruptions~\cite{belldigital}.

In modern \gls{ran} architectures—including the \gls{oran} framework—the threshold for ``real‑time'' performance varies by control‑loop function and application. Non‑real‑time \gls{ric} applications (\emph{rApps}) operate on timescales exceeding one second for long‑term optimization. Near‑real‑time functions (\emph{xApps}) execute within ten to one thousand milliseconds for time‑sensitive control~\cite{bonati2020intelligence}. The most stringent tier, distributed applications (\emph{dApps}), enables sub‑ten‑millisecond response times~\cite{doro_dapps_2022, lacava_dapps_2025}. Accordingly, digital‑twin systems must meet these diverse latency targets, with the strictest real‑time requirement—data acquisition and response—set below ten milliseconds for dApp‑level operations. To achieve a digital twin framework that is effective across all layers of the protocol stack, accurate and agile channel modeling is essential to precisely and swiftly characterize the radio signal propagation through a dynamic environment.
This constitutes the basis of a high-fidelity digital twinning system~\cite{belldigital}.

Channel modeling serves as the foundational layer upon which all digital twin capabilities are built. Without accurate propagation models, digital twins cannot reliably predict network behavior, optimize resource allocation, or trigger preemptive actions in the physical network. The challenge lies in achieving the dual requirements of high-accuracy and ultra-low-latency: channel models must be computed fast enough to support real-time decision making while maintaining sufficient fidelity to ensure reliable network operations. This creates a fundamental tension between computational complexity and model accuracy that existing approaches struggle to resolve.\\

\noindent\textbf{\textit{A Primer on Current Approaches to Channel Modeling.}}
Channel modeling for RF scenarios traditionally falls into three main categories: measurement-based models, statistical models, and deterministic methods (e.g., ray‑tracing). 
Measurement‑based approaches capture site‑specific phenomena with high accuracy but are costly, labor‑intensive, and quickly outdated in dynamic environments~\cite{zhao2020playback}.
%
Statistical channel models employ stochastic or deterministic mathematical equations to characterize wireless propagation~\cite{3gpp38901}. However, these models often fail to represent all scenarios or capture environmental intricacies. Their reliance on simplified environmental assumptions leads to prediction inaccuracies, particularly in site-specific scenarios~\cite{zemen_site-specific_2025}.
%
Ray tracing provides a deterministic means of modeling wireless channels by launching rays and simulating their interactions—reflection, diffraction, and transmission—with environmental geometries using material-specific parameters~\cite{fuschini_ray_2015}. Although it achieves higher site-specific accuracy than empirical models---especially in architecturally complex settings---it remains impractical for real‑time use. Even with GPU‑accelerated implementations, the computational burden of dynamic mobility modeling and scenario adaptation is prohibitive, and the exhaustive pre-computation generally demands extensive storage. Furthermore, its fidelity depends on highly detailed 3D maps with accurate material assignments and accurate antenna patterns, since electromagnetic responses vary noticeably across surfaces. Therefore, routinely, in a trade-off against computational burden, ray tracing remains an approximation—constrained by finite ray sampling and often incomplete modeling of propagation phenomena like diffraction, scattering, and reflection for all possible points. These factors limit ray tracing’s suitability for high‑fidelity channel modeling in digital twins.

Advanced propagation modeling attempts to find a balance between accuracy and computational efficiency. \gls{dl} excels in this regard by training on extensive propagation datasets to capture complex channel behaviors without explicit geometric simulation. 
\gls{ai}/\gls{ml} frameworks can automatically learn nonlinear relationships among environmental features, measurements, and spatial dynamics. Consequently, they can deliver high‑fidelity, real‑time channel estimates, generalize across varied scenarios, and continuously refine their estimations as new data arrive---all while sidestepping the computational burdens inherent in conventional approaches~\cite{aldossari_machine_2019, huang_big_2020}.

Radio (environment) maps provide a two‑dimensional representation of averaged statistics of channel characteristics (e.g., received signal power, interference power, power spectral density, delay spread, and channel gain) over a geographic region~\cite{el-friakh_crowdsourced_2018}. Unlike individual point-wise channel estimates, radio maps capture spatial relationships and large-scale propagation patterns, reflecting how neighboring locations influence one another. This inherent spatial structure makes radio maps a natural and effective output format for \gls{dl} models, which can leverage locality and spatial dependencies. For instance, \gls{cnn} architectures can process environmental inputs such as terrain and building layouts to generate channel predictions for an entire area in a single inference step, rather than predicting each point independently. This structured approach not only improves scalability and inference speed but also provides the spatial context essential for real-time digital twin applications.

This paper presents {\bf AIRMap, a deep-learning-based framework for real-time, high-fidelity radio map estimation. }
AIRMap is trained on the largest site-specific dataset of its kind, automatically generated through a scalable pipeline using ray-tracing simulations and 2D elevation data. 
Unlike prior models requiring multiple inputs, AIRMap leverages a single-channel elevation map to produce accurate channel predictions with sub-4 dB RMSE and millisecond-level inference latency. 
A lightweight calibration procedure using sparse field measurements significantly improves accuracy, reducing median error to 5\%. 
AIRMap is integrated into two state-of-the-art platforms--Sionna SYS and the Colosseum testbed--where it achieves near-zero error on spectral efficiency and block error rate metrics, validating its suitability for real-time digital twin applications across protocol layers.

The main contributions of this paper are as follows:
\begin{itemize}
    \item \textbf{Ray-Tracing Efficiency Analysis}: We systematically analyze the computational complexity-fidelity trade-offs in ray-tracing simulations, identifying optimal configurations that enable large-scale dataset generation while maintaining high fidelity for deep learning applications.
    
    \item \textbf{Large-Scale Radio Map Dataset}: We develop an automated pipeline to generate the largest site-specific radio map dataset to date, comprising 1.2M Boston-area samples with diverse propagation scenarios for robust neural network training.

    \item \textbf{Variable-Scale Coverage Modeling}: Each sample covers a square region with side lengths ranging from 500\,m to 3\,km, allowing the model to generalize across both local and wide-area propagation conditions.

    \item \textbf{Single-Input U-Net Model}: We design a novel neural network architecture that requires only 2D elevation maps as input, achieving sub-4\,dB RMSE with 4\,ms inference time, over $100\times$ faster than GPU-accelerated ray tracing.

    \item \textbf{Cross-Environment Generalization}: We validate AIRMap across four distinct propagation environments spanning flat urban, mountainous urban, flat rural, and mountainous rural regions, demonstrating strong generalization beyond the training environment.
    
    \item \textbf{Fine-tuning}: We develop a calibration framework that leverages large-scale simulated ray-tracing data for pretraining, followed by fine-tuning with a small subset of real-world measurements. This approach reduces median prediction error from over 50\% to approximately 5\%, effectively bridging the simulation-to-reality gap with minimal measurement overhead while preserving model generalization and accuracy.
    
    \item \textbf{Real-Time Testbed Integration}: We demonstrate the first sub-10ms radio map generation in operational wireless testbeds (Colosseum and Sionna SYS), enabling true real-time digital twin applications with near-zero system-level performance error.
\end{itemize}

\review{
While AIRMap demonstrates strong accuracy and real-time performance, several limitations highlight important directions for future research. In its current form, the model is trained at a fixed carrier frequency of 1 GHz with isotropic antenna patterns; extending the framework to other frequency bands or directional antenna configurations may therefore require recalibration. Nevertheless, the transfer-learning strategy introduced in this work suggests that such adaptation can be efficiently achieved using limited, targeted measurement data. Moreover, AIRMap currently focuses on predicting scalar path-gain radio maps. Extending the framework to richer channel representations—such as full \gls{cir}, delay spread, and angular characteristics—would further enhance the fidelity of wireless digital twins. Future work will also investigate model generalization in more challenging propagation environments, including indoor–outdoor transition regions, as well as the interpretability of the learned representations to better understand how environmental features influence propagation behavior. 
}

The remainder of this paper is organized as follows.
Section~\ref{sec:model_reality} evaluates conventional channel modeling techniques, focusing on ray tracing and its limitations through a real-world measurement campaign. 
We identify key factors—such as material properties and antenna patterns—that contribute to the gap between simulated and measured propagation data.
In Section~\ref{sec:dataset}, we present our automated pipeline for generating a large-scale dataset using Sionna RT~\cite{sionna}, detailing trade-offs between ray-tracing fidelity and computational efficiency. 
We also describe the dataset structure and its suitability for deep learning.
Section~\ref{sec:ai} introduces AIRMap, our U-Net-based \gls{ai} framework for real-time radio map estimation. 
We outline the model architecture, training approach, and a lightweight calibration pipeline using sparse measurements to enhance accuracy.
In Section~\ref{sec:digital_twin}, we demonstrate the integration of AIRMap into the Sionna SYS simulator and the Colosseum testbed, validating its accuracy and real-time performance in system-level simulations and channel emulation.
%
Finally, Section~\ref{sec:conclusion} concludes the paper and discusses future directions.
\review{
The source code for dataset generation is publicly available at \url{https://github.com/wineslab/sionna_data_generator}. 
}

\section{Practical Analysis of Ray-tracing-based Channel Models}
\label{sec:model_reality}

\begin{figure}[!bp]
    \centering
    \includegraphics[width=0.48\textwidth]{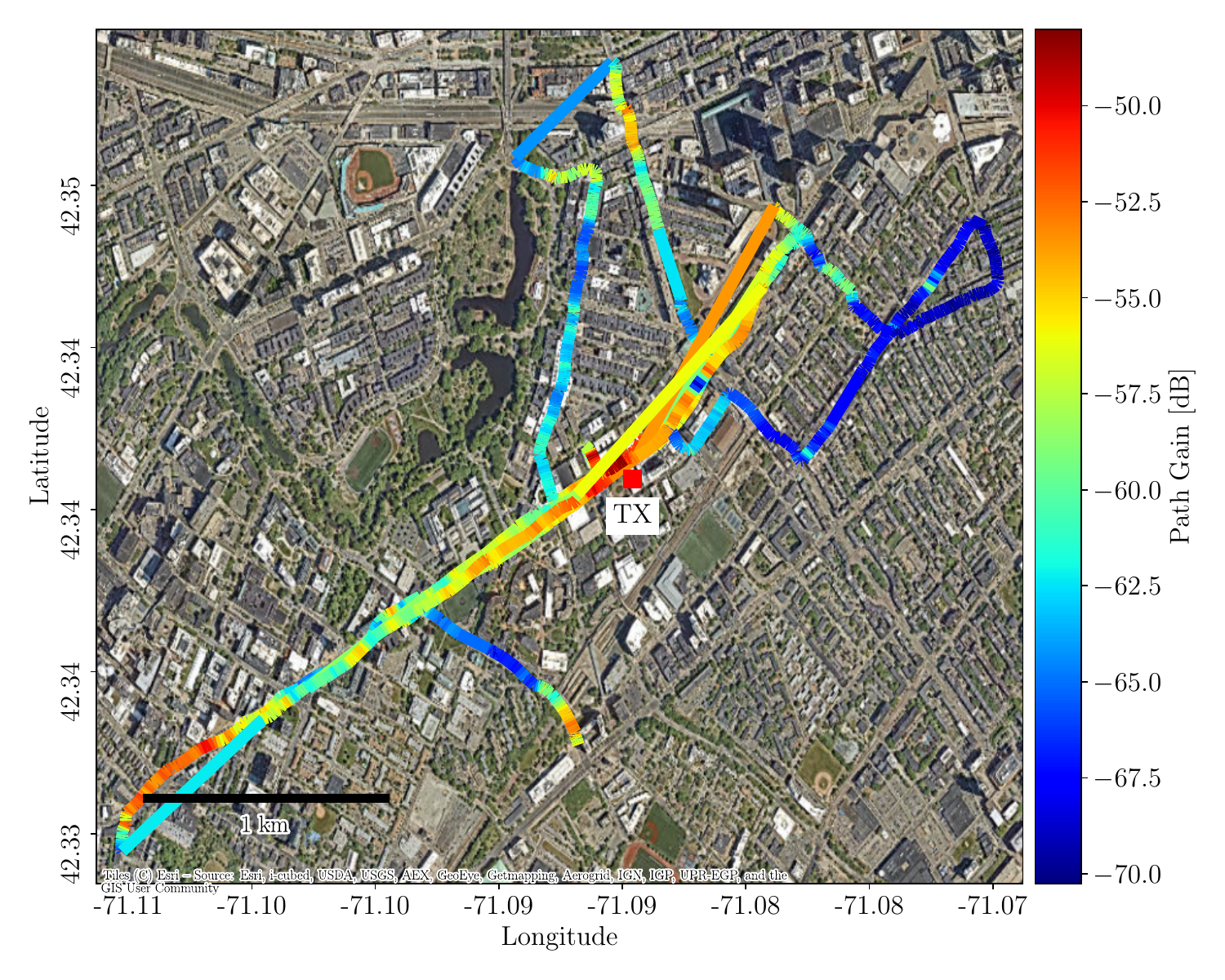}
    \caption{Measurement campaign map showing the \gls{tx} location and \gls{rx} route. The line color represents the path gain and coverage distribution.}
    \label{fig:meas_route}
\end{figure}


%
In this section, we analyze a measurement scenario to assess the impact of simulation parameters on the gap between ray-tracing predictions and real-world measurements. Since simulation data forms the foundation for training the \gls{dl} model, we propose a calibration approach that uses a small set of field measurements to fine-tune the model. This enables the data-driven model to surpass ray tracing in accuracy by correcting for systematic simulation errors.

\noindent\textbf{\textit{Measurement Campaign}}. We design a measurement scenario to capture channel characteristics at 933 MHz within a section of the Northeastern University campus. In this setup, the \gls{tx} remains stationary on the rooftop of Dodge Hall, while the \gls{rx} is mobile, following the route depicted in Fig.~\ref{fig:meas_route}.

For the \gls{tx} side, we used an USRP X410 as the \gls{ru}, followed by a Minicircuits ZHL-1000-3W+ amplifier, and a Pasternack PE51OM1014 antenna. On the \gls{rx} side, we employed the same antenna as used in the \gls{tx} side. The \gls{rx} \gls{ru} was the Viavi Solutions T/Rx Software Defined Transceiver~\cite{viavi_solutions}.

%

%
For channel measurements, we utilized the maximum available bandwidth of 2\,MHz at a center frequency of 933\,MHz. Transmission was conducted in bursts of 500 consecutive GLFSR-14 codewords every 0.1 seconds. All losses and gains from cables, amplifiers, antennas, and other hardware components were removed by characterizing the entire setup in an anechoic chamber under identical conditions. A summary of the measurement setup is provided in Table~\ref{tab:meas-setup}.

\noindent\textbf{\textit{Path Gain Computation from CIR}}.  
We compute the path gain directly from the measured \gls{cir}.
\begin{equation}
h(t, \tau) = \sum_{i=1}^{N} \alpha_i(t) \delta\left(\tau - \tau_i(t)\right)
\label{eqn:cir}
\end{equation}
where $\alpha_i(t)$ is the complex amplitude and $\tau_i(t)$ is the delay of the $i$-th path at time $t$. 

The total path gain at each instant can then be computed by integrating the squared magnitude of the CIR.
\begin{equation}
\begin{aligned}
    P_{\text{RX}}(t)
    &= 10 \log_{10} \left( \sum_{i} \left|\alpha_i(t)\right|^2 \right)
\label{eqn:p_rx}
\end{aligned}
\end{equation}

In our mobile measurement scenario, the receiver's location changes over time; thus, we can associate each measurement with the receiver position $\boldsymbol{q}_{\mathrm{RX}}(t)$. Under this setting, the time-indexed path gain $\text{P}(t)$ can be equivalently expressed as a location-dependent function $\text{P}(\boldsymbol{q}_{\mathrm{RX}})$, allowing us to map the measured data directly into spatially indexed radio maps.

\begin{table}[!bp]
    \centering
    \caption{Measurement Setup and Equipment}
    \label{tab:meas-setup}
    \begin{tabular}{p{2.5cm}p{5cm}}
    \toprule
    \textbf{\textit{\gls{tx}}} & \\ 
    \midrule
    \gls{ru} & Ettus USRP X410 \\ 
    Amplifier & Minicircuits ZHL-1000-3W+ (38 dB) \\ 
    Antenna & Pasternack PE51OM1014 (6 dBi) \\ 
    Location & 42°20'25"N 71°05'16"W \\
    \toprule
    \textbf{\textit{\gls{rx}}} & \\ 
    \midrule
    \gls{ru} & T/RX provided by VIAVI Solutions \\
    Antenna & Pasternack PE51OM1014 (6 dBi) \\ 
    Location & Mobile (see Fig.~\ref{fig:meas_route}) \\
    \toprule
    \textbf{\textit{Meas. Details}} & \\
    \midrule
    Frequency & 933 MHz \\ 
    Bandwidth & 2 MHz \\ 
    Codeword & GLFSR-14 \\ 
    Synchronization & GPS clock for both \gls{tx} and \gls{rx} \\ 
    \bottomrule
    \end{tabular}
\end{table}

\begin{figure}
    \centering
    \begin{subfigure}[b]{0.45\textwidth}
        \centering
        \includegraphics[width=\textwidth]{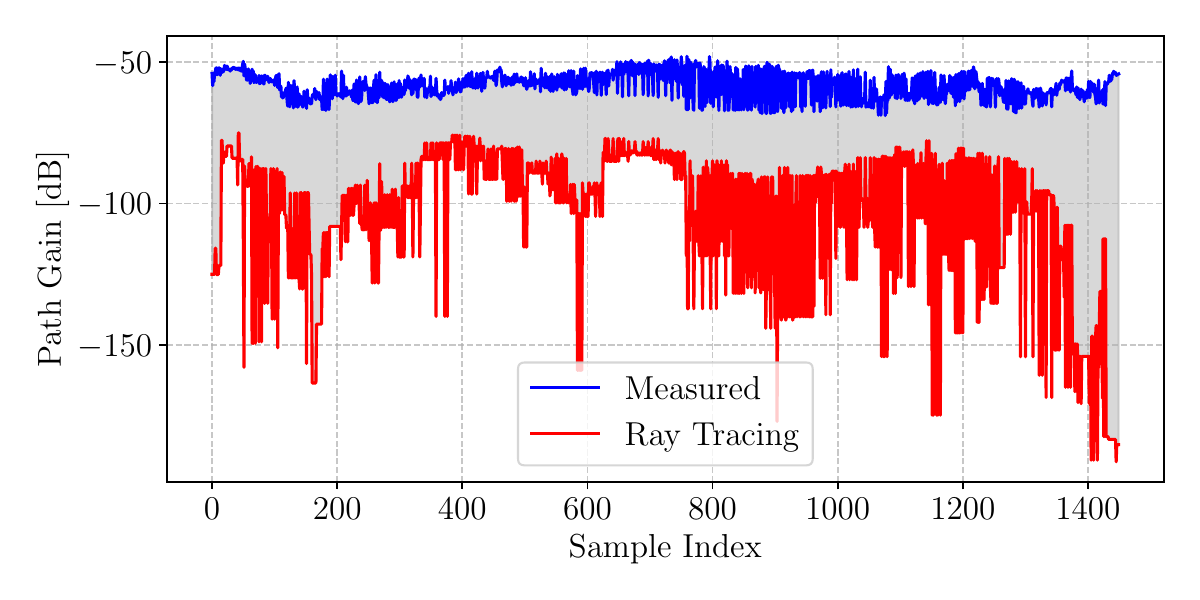}
        \caption{Comparison of measured and ray‑traced path gain values across locations. Maximum normalized correlation is 0.3967 at zero lag.}
        \label{fig:pg_comp}
    \end{subfigure}
    \begin{subfigure}[b]{0.45\textwidth}
        \centering
        \includegraphics[width=\textwidth]{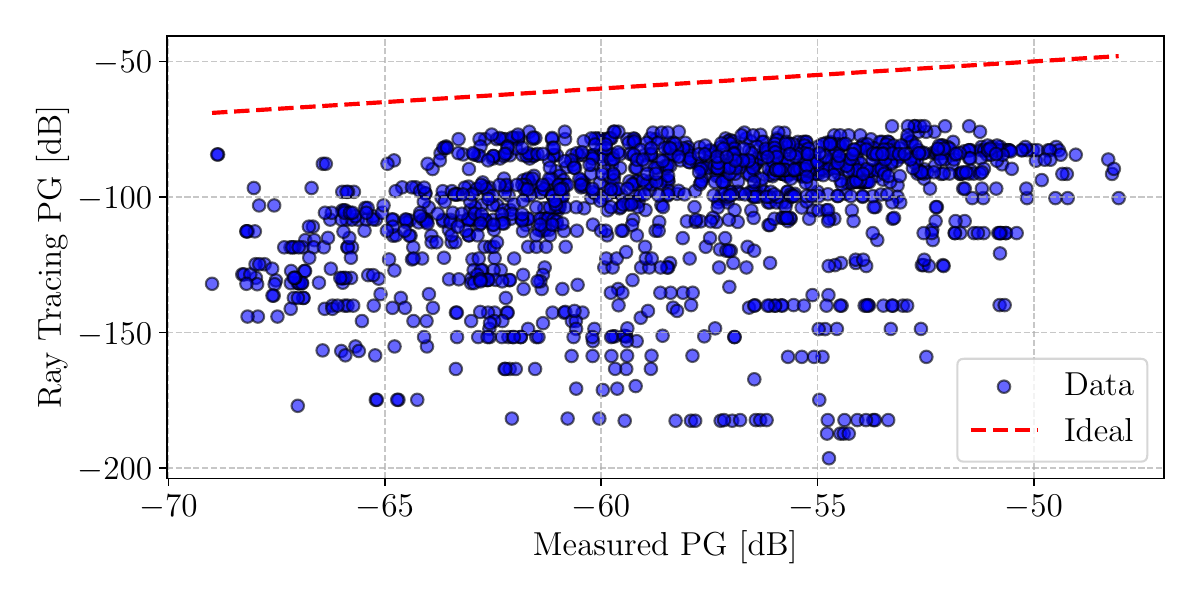}
        \caption{Scatter plot of the ray tracing results vs. measured values.}
        \label{fig:pg_scatter}
    \end{subfigure}
    \caption{Evaluation of ray tracing performance by comparing with actual measurement data.}
    \label{fig:model_evaluation}
\end{figure}

In Fig.~\ref{fig:model_evaluation}, we compare ray‑tracing (Sionna) results—generated using a maximum‑detail configuration on the same locations as our measurements—to the empirical data. The 3D environment model was sourced from the Boston Planning Department~\cite{testolina_boston_2024}, with all surfaces assigned concrete material properties and terrain modeled as dry ground. As shown in Fig.~\ref{fig:pg_comp}, ray tracing and measured path gains exhibit substantially different levels. The scatter plot in Fig.~\ref{fig:pg_scatter} further illustrates that these differences persist across all sample points, deviating from the ideal linear relationship. This discrepancy highlights the limitations of ray tracing’s approximations and its inability to fully capture real‑world propagation.  

\begin{figure}[!ht]
    \centering
    \includegraphics[width=0.98\linewidth]{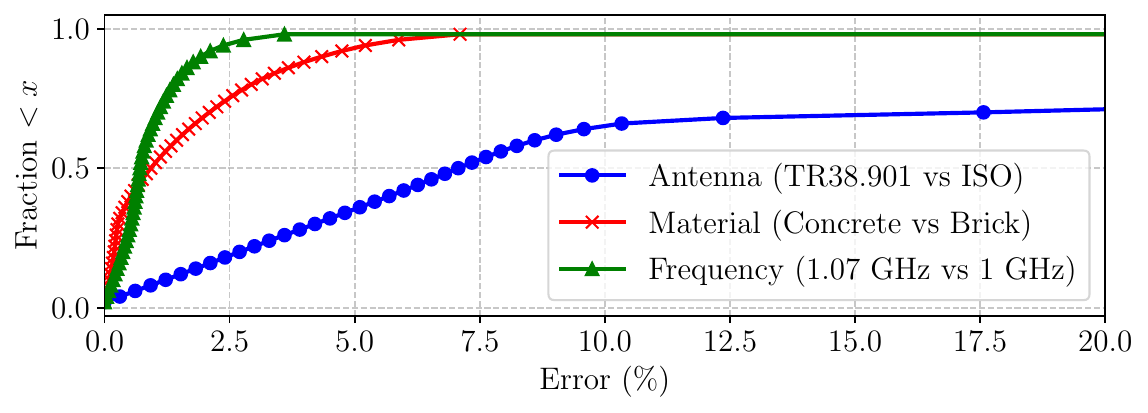}
    \caption{\review{\gls{ecdf} of path gain error caused by material, antenna-pattern, and frequency misconfigurations in ray tracing.}}
    \label{fig:misconfig}
\end{figure}

\review{
Calibration techniques that adjust material properties can improve ray-tracing fidelity; however, such methods are typically constrained to small-scale indoor environments and do not generalize effectively to large outdoor deployments~\cite{kanhere2023calibration,hoydis2023learning,jemai2009calibration}. To systematically quantify the impact of simulation parameter misconfigurations on path-gain prediction accuracy—particularly in scenarios where detailed environmental parameters are unavailable—we conducted a comprehensive sensitivity analysis using Sionna RT. A total of 100 randomized scenarios were simulated with diffraction enabled and a maximum interaction depth of 20. Three sources of misconfiguration were evaluated independently: (i) material properties, by assigning ITU brick to all surfaces while assuming concrete as the ground-truth material; (ii) antenna radiation patterns, by comparing the 3GPP TR\,38.901 specification against an isotropic radiator; and (iii) carrier frequency offset, by introducing a 70\,MHz deviation from the nominal 1\,GHz operating frequency. The results, summarized in Fig.~\ref{fig:misconfig}, demonstrate the degradation in path-gain accuracy attributable to each misconfiguration source, underscoring the sensitivity of ray-tracing simulations to parameter assumptions and motivating the need for data-driven calibration approaches.
}

\begin{figure}
    \centering
    \includegraphics[width=0.5\textwidth]{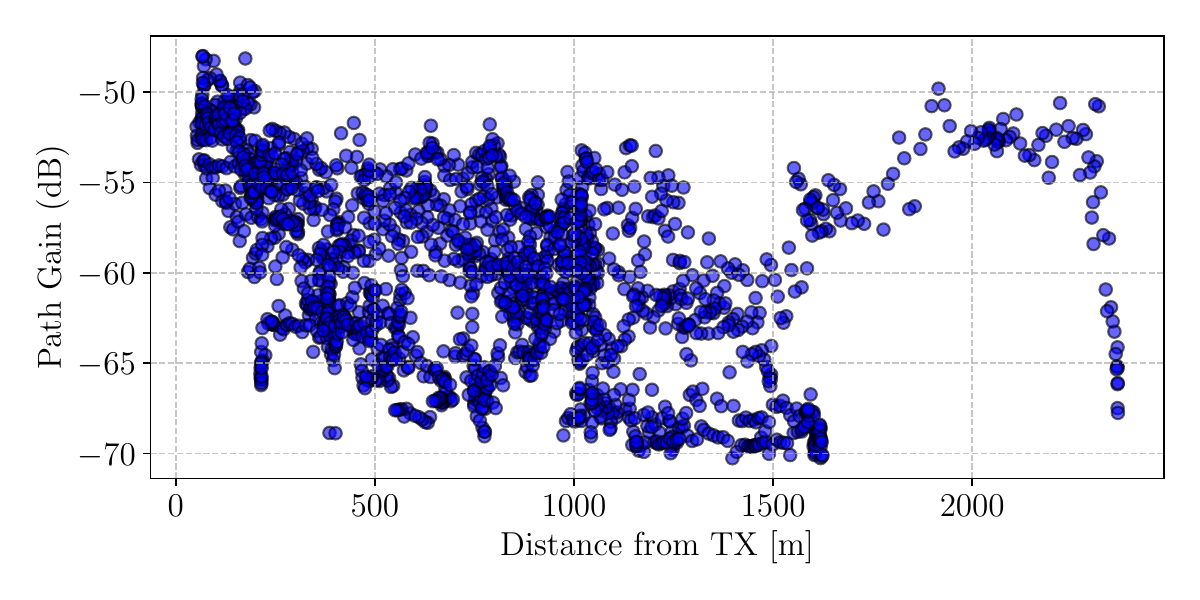}
    \caption{Measurement results ordered by distance from the \gls{tx}, highlighting the difficulty of fitting traditional empirical models to real-world data.}
    \label{fig:pg_distance}
\end{figure}

In Fig.~\ref{fig:pg_distance}, we plot the measured path gain values ordered by distance from the \gls{tx}. The highly irregular urban geometry prevents a simple radial approximation, as signal propagation is neither symmetric nor uniform. Consequently, site‑specific inputs—such as detailed environmental geometry—are essential. Traditional statistical models, which are not tailored to a particular site, fail to capture the complex, nonlinear propagation effects present in urban areas and therefore lack sufficient accuracy. It can be observed in Fig.~\ref{fig:pg_distance} that it is not trivial to find a low-degree polynomial that fits the variations in the path gain at different distances from the transmitter.

\section{Dataset}
\label{sec:dataset}
\subsection{Efficient Generation of High-Fidelity Radio Maps}

\begin{figure}[!ht]
    \centering
    \includegraphics[width=0.45\textwidth]{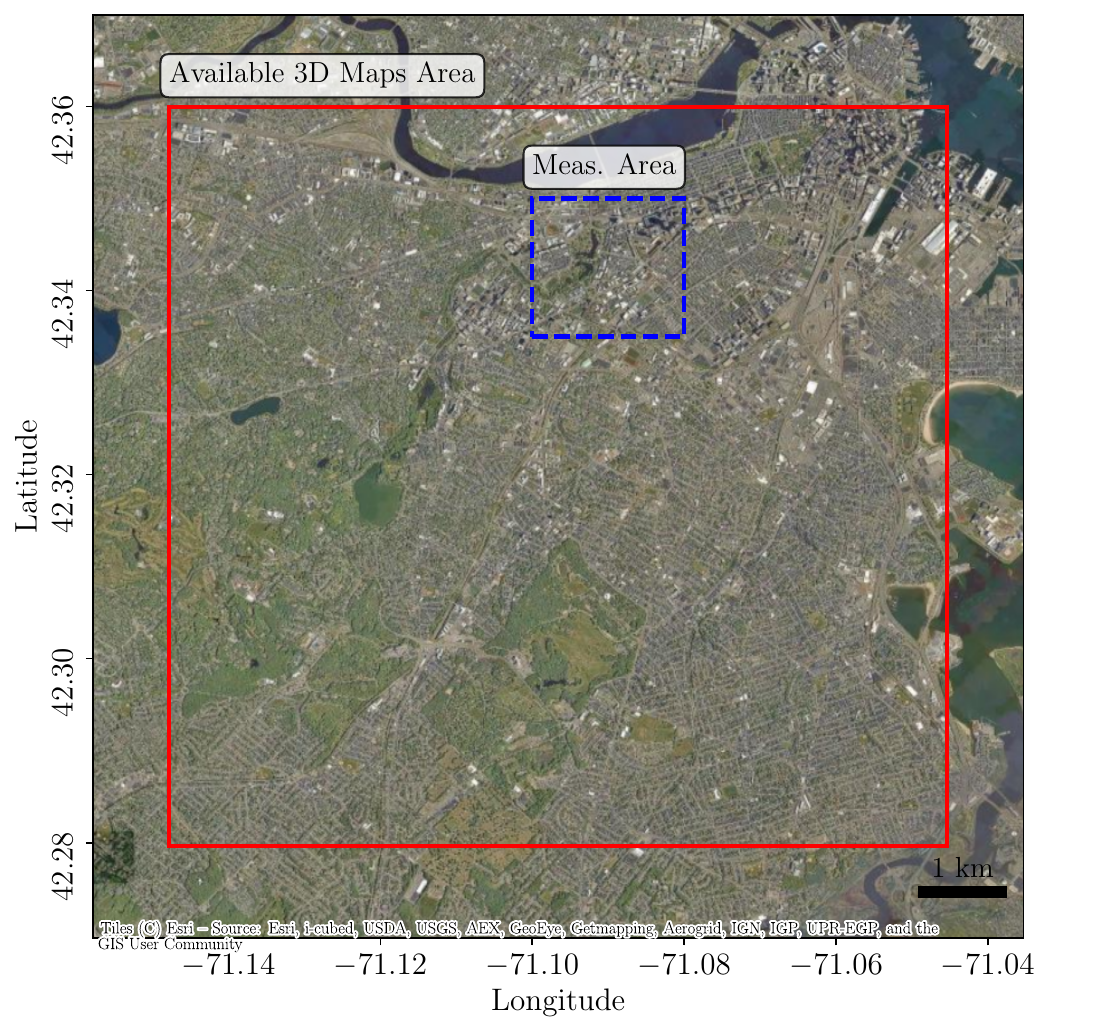}
    \caption{Geographic study area in Boston. The red-outlined region marks the environment used for dataset generation, while the blue-highlighted region corresponds to the area used in the measurement campaign shown in Fig.~\ref{fig:meas_route}.}
    \label{fig:boston_map}
\end{figure}

\begin{figure}[!htp]
    \centering
    \includegraphics[width=0.49\textwidth]{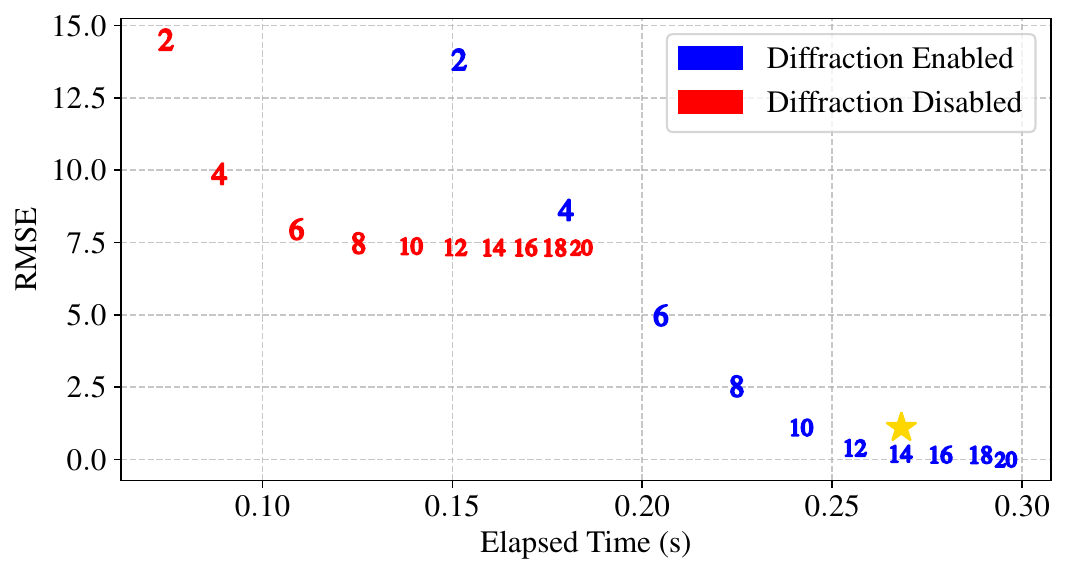}
    \caption{\gls{rmse} versus runtime for varying ray‑tracing configurations in Sionna RT (1.2.1). Marker labels denote maximum path depth (reflections).}
    \label{fig:rt_config}
\end{figure}

\begin{figure*}
    \centering
    \includegraphics[width=0.95\linewidth]{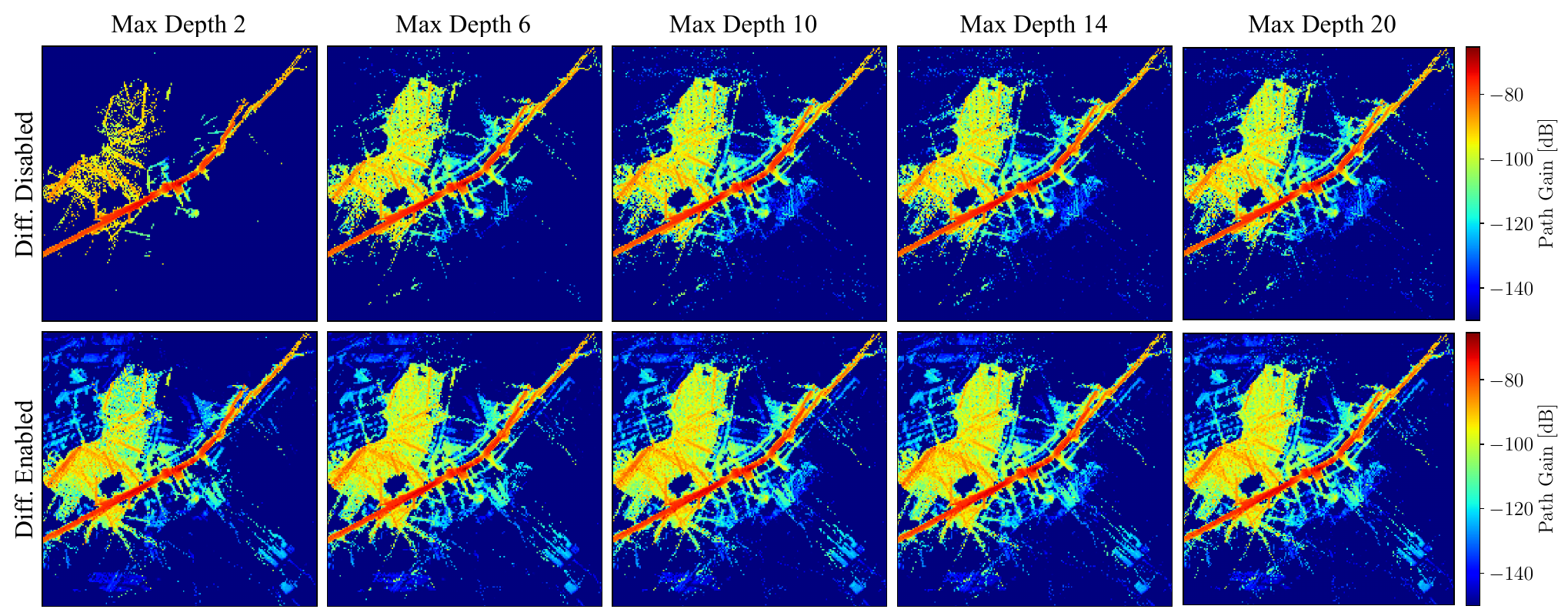}
    \caption{Evolution of radio maps with increasing ray-tracing depth. Each column corresponds to a different maximum path depth. Rows compare configurations with diffraction disabled (top) and enabled (bottom).}
    \label{fig:rt_config_evolution}
\end{figure*}

Ray-tracing offers high-fidelity channel modeling but remains computationally expensive, especially when creating the large-scale datasets needed for training \gls{dl} models. To enable scalable dataset generation, we explore the trade-off between simulation accuracy and runtime by varying ray-tracing configurations.

Among various ray-tracing tools, we selected Sionna RT~\cite{sionna} (version 1.2.1) for its ease of automation, native Python interface, and GPU acceleration—features that streamline large-scale dataset generation. Built on top of Mitsuba 3, a differentiable rendering system powered by the just-in-time compiler Dr.Jit, Sionna RT offers a flexible and efficient framework for radio propagation modeling, making it particularly well-suited for AI-driven wireless system design.

We use the Boston area shown in Fig.~\ref{fig:boston_map} as a representative urban environment and randomly select 100 \gls{tx} locations placed outside buildings. For each ray-tracing configuration, we generate path gain radio maps and evaluate accuracy using the \gls{rmse} metric. The configuration with diffraction enabled and a maximum depth of 20 is used as the ground truth (maximum configuration). All other configurations are compared against this baseline to identify the most efficient setup that balances accuracy and simulation time. All simulations are conducted using Sionna RT on an NVIDIA L40S GPU. Reported runtimes exclude scene initialization and geometry loading to focus solely on ray-tracing execution.

Fig.~\ref{fig:rt_config} shows the simulation runtime for 100 scenarios (each repeated five times) across different ray-tracing configurations. As expected, enabling diffraction and increasing the maximum path depth both lead to longer runtimes. However, when diffraction is enabled, the \gls{rmse} levels off beyond a path depth of 14, indicating diminishing returns in accuracy. Fig.~\ref{fig:rt_config_evolution} illustrates how varying these parameters affects the resulting path-gain radio maps in a representative scenario. Based on this analysis, we identify a “sweet spot” configuration—diffraction enabled with a path depth of 14—that balances fidelity and efficiency, requiring approximately 0.265 seconds to generate each radio map.

\subsection{Automated Generation of Large-Scale Datasets}

For all \gls{tx} locations within the Boston area (see Fig.~\ref{fig:boston_map}), we load the scene geometry from the BostonTwin model~\cite{testolina_boston_2024} into Sionna RT with a carrier frequency of 1\,GHz. Both transmitter and receiver use vertically polarized isotropic antennas, and the transmit power is fixed at 44~dBm to match typical urban macrocell deployments~\cite{deruyck_power_2014}. We then generate path‑gain radio maps for each \gls{tx} position.  

\review{
For training data, we randomly select 240,000 valid TX locations within the rectangle defined by latitudes 42.2796°–42.3599°N and longitudes 71.1478°–71.0453°W—a region spanning approximately $100~km^{2}$ of the greater Boston area—excluding points inside building footprints. Each sample comprises a pair of uniformly cropped, rasterized maps: (1) the path-gain radio map and (2) the corresponding 2D building elevation map, with the TX centered in both. To introduce variability in coverage area, the map extent for each sample is drawn uniformly between 500 m and 3 km, corresponding to individual sample areas ranging from 0.25 km² to 9 km². This pipeline can ingest any 3D urban dataset for which detailed geometry is available, enabling seamless generalization to new locations.
}

\review{
To mitigate overfitting and enhance model robustness, we apply geometric data augmentation: each sample is rotated by 90°, 180°, and 270°, and flipped horizontally and vertically. Because the TX is always centered in the input image, these transformations preserve the relative spatial relationship between the transmitter and surrounding geometry—only the orientation changes. This is equivalent to rotating the physical scenario while keeping the TX-RX geometry intact, ensuring that each augmented sample represents a physically valid propagation scenario. This augmentation strategy serves two purposes: (1) it increases dataset diversity to prevent overfitting, and (2) it enforces rotational invariance, ensuring the model learns that propagation physics are independent of cardinal direction. This expands the dataset to 1.2M samples, offering ample diversity for training deep-learning models. As a result, to our knowledge, this represents the largest radio-map dataset, with effectively unlimited extensibility in both geographic scope and sample count.
}

\section{DL-Based Estimation of Radio Maps}
\label{sec:ai}


%

\begin{sidewaystable}[!htbp]
\centering
\caption{Applications of Deep Learning in Channel Parameter Prediction}
\label{tab:dl_works}
\resizebox{\textwidth}{!}{
\begin{tabular}{lllp{3.5cm}lllp{3.7cm}}
\toprule
\textbf{Ref.} & \textbf{\gls{ai}/\gls{ml} Methods} & \textbf{Inputs} & \textbf{Predicted Parameters} & \textbf{Frequency} & \textbf{Environment} & \textbf{Data Source} & \textbf{Best Performance} \\
\midrule
\cite{IKI19} & \gls{cnn} + MLP & Distance map, building height & Path loss & 2\,GHz & Urban macrocell & Ray-tracing & RMSE = 9.94\,dB \\[0.5ex]

\cite{SGS20feature} & \gls{cnn}, MLP & Building height map & Path loss & 900\,MHz & 146 urban environments & Ray-tracing & RMSE = 4.42\,dB \\[0.5ex]

\cite{HNI20} & \gls{cnn} + MLP & Aerial image, building info, distance, angle, system parameters & RSS & 2\,GHz & Urban (Tokyo) & Measurements & RMSE = 3.96\,dB \\[0.5ex]

\cite{TZC20} & Correctional \gls{cnn} + MLP & Distance, relative geodetic distance, Rx coordinates, satellite image & RSRP & 0.811, 2.63\,GHz & Urban & Measurement, Ray-tracing & RMSE = 4.3\,dB @ 0.811\,GHz, RMSE = 4.2\,dB @ 2.63\,GHz \\[0.5ex]

\cite{TSW20} & Correctional \gls{cnn} + MLP & Delta lat/long, image of local surroundings, link budget rough estimate & RSS (RSSP) & 0.811, 2.63\,GHz & Urban, suburban, highway & Measurement & RMSE = 4.7\,dB @ 0.811\,GHz, RMSE = 4.1\,dB @ 2.63\,GHz, RMSE = 6.3\,dB across environments \\[0.5ex]

\cite{NCK22} & Correctional \gls{cnn} + MLP & Distance, relative geodetic distance, Rx locations, satellite image & Path loss & 3.5\,GHz & Urban & Simulation + Measurement & RMSE = 4.46\,dB \\[0.5ex]

\cite{IIN22} & \gls{cnn} + MLP & Spatial map, Tx-Rx spatial map, FSPL rough estimate map, system parameters & RSS & LTE 2.1\,GHz & Urban (Tokyo) & Measurement & MSE = 8.07 \\[0.5ex]

\cite{DCV22} & \gls{cnn} \gls{ae} & Spatial map (LiDAR + CADMAPPER) & Path loss & 28\,GHz & Urban canyon & Measurements & RMSE = 4.8\,dB \\[0.5ex]

\cite{OMC22} & \gls{cnn} \gls{ae} + U-Net & Spatial map, distance & RSS map & - & 5 different scenarios & Ray-tracing & MAE = 2.2\,dB (avg.) \\[0.5ex]

\cite{RCP21} & U-Net & Terrain height, building height, foliage height, LOS, BS height & Path loss & 28\,GHz & - & Ray-tracing & RMSE = 5.8\,dB \\[0.5ex]

\cite{LYK21} & U-Net & City map, Tx location, some path loss measurements & Path loss & 5.9\,GHz & Urban & Ray-tracing & - \\[0.5ex]

\cite{BCQ22} & U-Net & Permittivity, conductivity, distance, FSPL, spatial map & Path loss heat map & 0.433, 2, 3.7\,GHz & Indoor & Ray-tracing & RMSE = 4.21\,dB \\[0.5ex]

\cite{LSS22} & U-Net & Spatial map, Tx location & RSS map & 2.5\,GHz & Campus & Ray-tracing & NMSE = 0.0006752 \\[0.5ex]

\cite{QBS22} & U-Net & 3-channel image (building height, Tx location Gaussian kernel, distance) & Path loss & 30\,GHz & Urban & Ray-tracing & RMSE = 5.78\,dB \\ 
\bottomrule
\end{tabular}}
\end{sidewaystable}

\begin{sidewaystable}[!htbp]
\centering
\caption{Applications of Conventional Machine Learning in Channel Parameter Prediction}
\label{tab:ml_works}
\resizebox{\textwidth}{!}{
\begin{tabular}{lllp{3cm}lllp{4cm}}
\toprule
\textbf{Ref.} & \textbf{\gls{ai}/\gls{ml} Methods} & \textbf{Inputs} & \textbf{Predicted Parameters} & \textbf{Frequency} & \textbf{Environment} & \textbf{Data Source} & \textbf{Best Performance} \\
\midrule
\cite{OZW17} & Random Forest, AdaBoost, \gls{knn}, NN & Distance, path angle, vegetation, terrain complexity & RSSI & 2.4\,GHz & Suburban (ground-to-ground) & Measurement & MAE = 3.72 (Random Forest) \\[0.5ex]

\cite{ZWY18} & Random Forest, \gls{knn} & Path visibility, distance, elevation angle, Tx/Rx altitude & Path loss & 2.4\,GHz & Urban (air-to-air) & Ray-tracing & MAE = 2.27 (Random Forest) \\[0.5ex]

\cite{YZH19} & Random Forest, \gls{knn} & Rx coordinates, building stats, elevation angle & Path loss, delay spread & 2.4, 5.8, 28, 37\,GHz & Urban (air-to-ground) & Ray-tracing & RMSE = 1.64 (Random Forest) \\[0.5ex]

\cite{SGS20} & XGBoost, Random Forest & 23 features (LOS path, Rx area, Tx/Rx positions) & Path loss & 900, 1800\,MHz & Outdoor & Ray-tracing & MAE = 3.17 (XGBoost) \\[0.5ex]

\cite{SGL18} & \gls{rbf} & Snapshot index, distance & Path loss, shadow fading, DoA & 26\,GHz & Outdoor microcell & Measurement & RMSE = 4.52 \\[0.5ex]

\cite{ZDG19} & \gls{rbf} & Snapshot index, distance & Path loss, shadow fading, small-scale parameters & 26, 28\,GHz & Outdoor microcell, indoor & Measurement & RMSE = 0.44 \\[0.5ex]

\cite{HWB20} & \gls{rbf}, \gls{fc} & Tx/Rx coords, distance, carrier frequency & Received power, delay spread, angle spreads & 11–60\,GHz & Indoor & Measurement, GBSM model & RMSE = 2.63 (\gls{fc}) \\[0.5ex]

\cite{ZDL22} & \gls{mlp} & Distance, relative height, terrain, building occlusion & RSRP & 2.5\,GHz & Urban, dense urban & Measurement & MAE $<$ 5 \\
\bottomrule
\end{tabular}}
\end{sidewaystable}

\subsection{Related Work on AI/ML-based Channel Models}

In \gls{ai}/\gls{ml}-based channel modeling, a mapping function is established between the wireless environment and its corresponding channel properties, enabling accurate and realistic channel parameter representation. Over the past few years, numerous studies have explored and developed \gls{ai}/\gls{ml}-based channel models~\cite{seretis2021overview, huang2022part2, DCV22}.

Early studies~\cite{OZW17, ZWY18, YZH19, WHG19} primarily employed conventional machine learning techniques—such as Random Forest, \gls{knn}, and \gls{svm}—for channel path loss prediction, with Random Forest generally achieving the best performance. Seretis et al~\cite{SGS20feature} later compared Random Forest with XGBoost, integrating extensive feature engineering to reduce model complexity. Their results indicated that XGBoost yielded higher prediction accuracy. A detailed summary of these works is presented in Table~\ref{tab:ml_works}.

Subsequent efforts shifted towards shallow neural network approaches. For instance,~\cite{SGL18, ZDG19, HWB20} investigated techniques such as \gls{rbf} neural networks to predict time-varying path loss, shadow fading, and small-scale channel characteristics, including the number of propagation paths and angular statistics, particularly within \gls{gbsm}. Huang et al.~\cite{HWB20} compared the performance of \gls{rbf} networks with \gls{mlp} neural networks, reporting that MLP slightly outperformed \gls{rbf} by a fraction of a decibel.

Further, Zhang et al.~\cite{ZDL22} utilized an \gls{mlp} neural network with additional environmental features, including terrain type and building occlusion, to predict the signal strength coverage within an urban scenario. Their model demonstrated superior accuracy compared to traditional propagation models.







%
The introduction of \glspl{cnn} and their ability to learn spatial correlations in images led to a new generation of models that utilize images as input to neural networks. These approaches involve constructing images from propagation-related features, which \glspl{cnn} analyze to extract spatial patterns, allowing the model to perform automated feature engineering and improve prediction accuracy. 

Examples of image-based environmental representations include distance maps and building maps, as proposed by Imai et al.~\cite{IKI19}, as well as low-resolution building height information used in~\cite{SGS20}. These \gls{cnn} models are typically followed by \gls{fc} layers for regression tasks, predicting channel parameters. Additionally, some studies have incorporated system parameters—such as frequency and antenna tilt—into the \gls{fc} layers as extra features, as these parameters may not be easily represented within an image format. 

Beyond structured maps, satellite and aerial images have also been integrated as additional input channels to \gls{cnn} models~\cite{HNI20, TZC20}. The impact of various environmental map images was explored in~\cite{HNI20}, while~\cite{IIN22} examined how different image sizes and spatial map construction methods between \gls{tx} and \gls{rx} affect prediction accuracy. Their findings indicate that a building occupancy rate map is more effective than aerial imagery and that including both \gls{rx} and \gls{tx} image data achieves similar performance to adding system parameters.

Hybrid approaches have also been explored, where a physics-based model is used to generate rough parameter estimates, which are then refined using a neural network. This correction-based method, proposed in~\cite{NCK22, TZC20}, improves prediction performance by leveraging domain knowledge in combination with deep learning. In these studies, rough estimates were integrated into the \gls{fc} regression layer, whereas in~\cite{IIN22}, a heat map of rough estimates—constructed using a free-space model—was incorporated as an additional input channel for the \gls{cnn} model.

Recent studies have focused on leveraging richer environmental information as input images to predict channel parameter heat maps in a single step. Works such as~\cite{RCP21, LYK21, BCQ22, LSS22, QBS22, OMC22, DCV22} have explored \gls{cnn}-based autoencoders and U-Net architectures for fast and accurate predictions. While U-Net enhances prediction accuracy, it comes at the cost of increased computational complexity. A summary of \gls{dl}-based methods for channel parameter prediction is presented in Table~\ref{tab:dl_works}.

\begin{table}[!t]
\centering
\caption{Network Architecture of the Modified PMNet}
\label{tab:net_arch}
{\footnotesize
\begin{tabular}{@{}clc@{}}
\toprule
\multicolumn{3}{c}{\textbf{Encoder}} \\ 
\midrule
\textbf{\#} & \textbf{Layer Type} & \textbf{Output Size} \\
\midrule
\multicolumn{3}{l}{\textbf{Input:} Image ($1 \times 200 \times 200$)} \\
1 ($\downarrow$) & Conv2d (64 ch), MaxPool2d & $64 \times 100 \times 100$ \\
2 & ResLayer (3 blocks) & $256 \times 100 \times 100$ \\
3 ($\downarrow$) & ResLayer (3 blocks) & $512 \times 50 \times 50$ \\
4 ($\downarrow$) & ResLayer (27 blocks) & $512 \times 25 \times 25$ \\
5 & ResLayer (3 blocks) & $1024 \times 25 \times 25$ \\
6 & ASPP + Conv2d (fc1) & $1024 \times 25 \times 25$ \\
\midrule
\multicolumn{3}{c}{\textbf{Decoder}} \\ 
\midrule
\textbf{\#} & \textbf{Layer Type} & \textbf{Output Size} \\
\midrule
\multicolumn{3}{l}{\textbf{Output:} Image ($1 \times 200 \times 200$)} \\
1 ($\uparrow$) & Conv2d & $512 \times 25 \times 25$ \\
2 ($\uparrow$) & ConvTranspose2d & $512 \times 50 \times 50$ \\
3 ($\uparrow$) & ConvTranspose2d & $256 \times 100 \times 100$ \\
4 & Conv2d & $256 \times 100 \times 100$ \\
5 & Conv2d & $128 \times 100 \times 100$ \\
6 & Conv2d + final head & $1 \times 200 \times 200$ \\
\bottomrule
\end{tabular}
}
\end{table}

\begin{figure}[!htp]
    \includegraphics[width=0.45\textwidth]{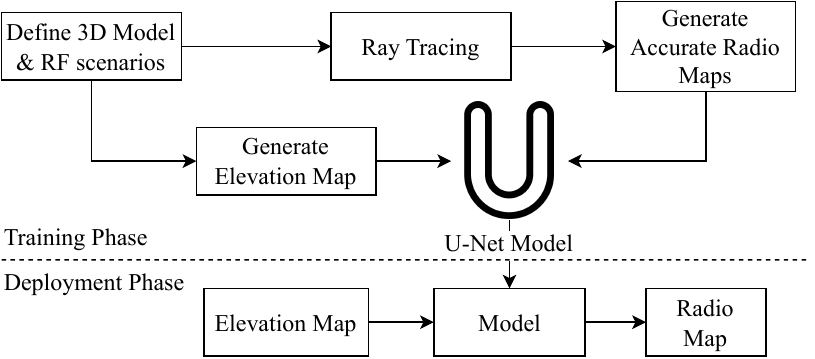}
    \caption{\gls{dl}-based model for radio map generation.}
    \label{fig:ai_emulation_process}
\end{figure}

\begin{figure}[!htp]
    \centering
    \includegraphics[width=0.45\textwidth]{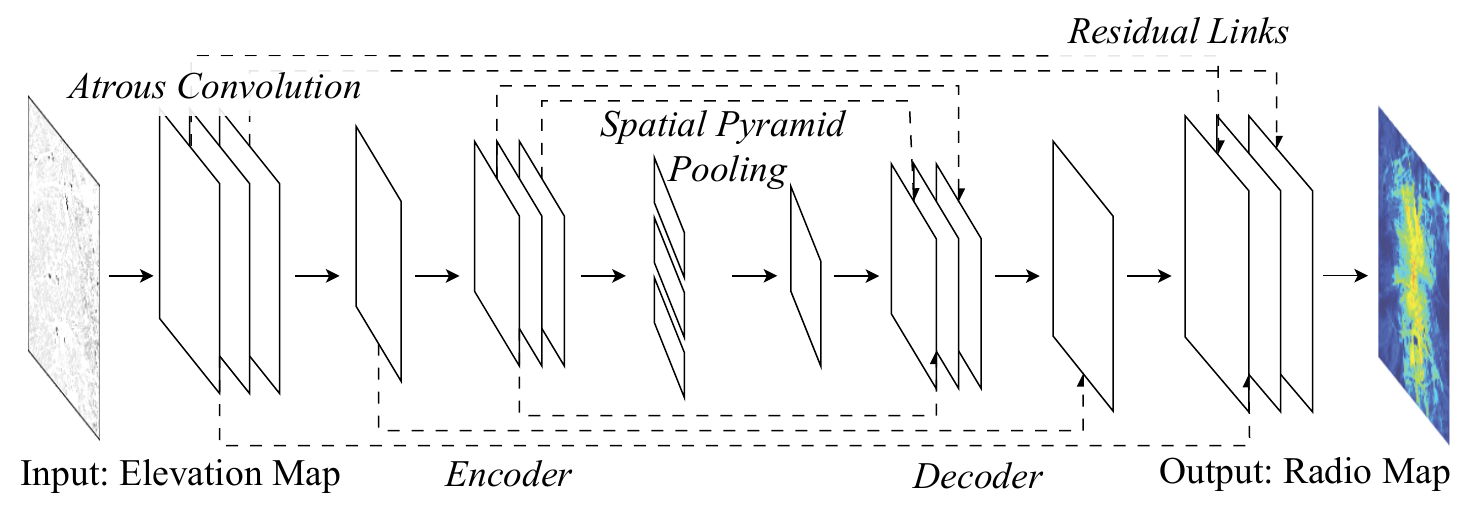}
    \caption{Proposed U-Net architecture for channel modeling.}
    \label{fig:net_arch}
\end{figure}

\begin{figure*}[!hb]
    \centering
    \includegraphics[width=0.95\linewidth]{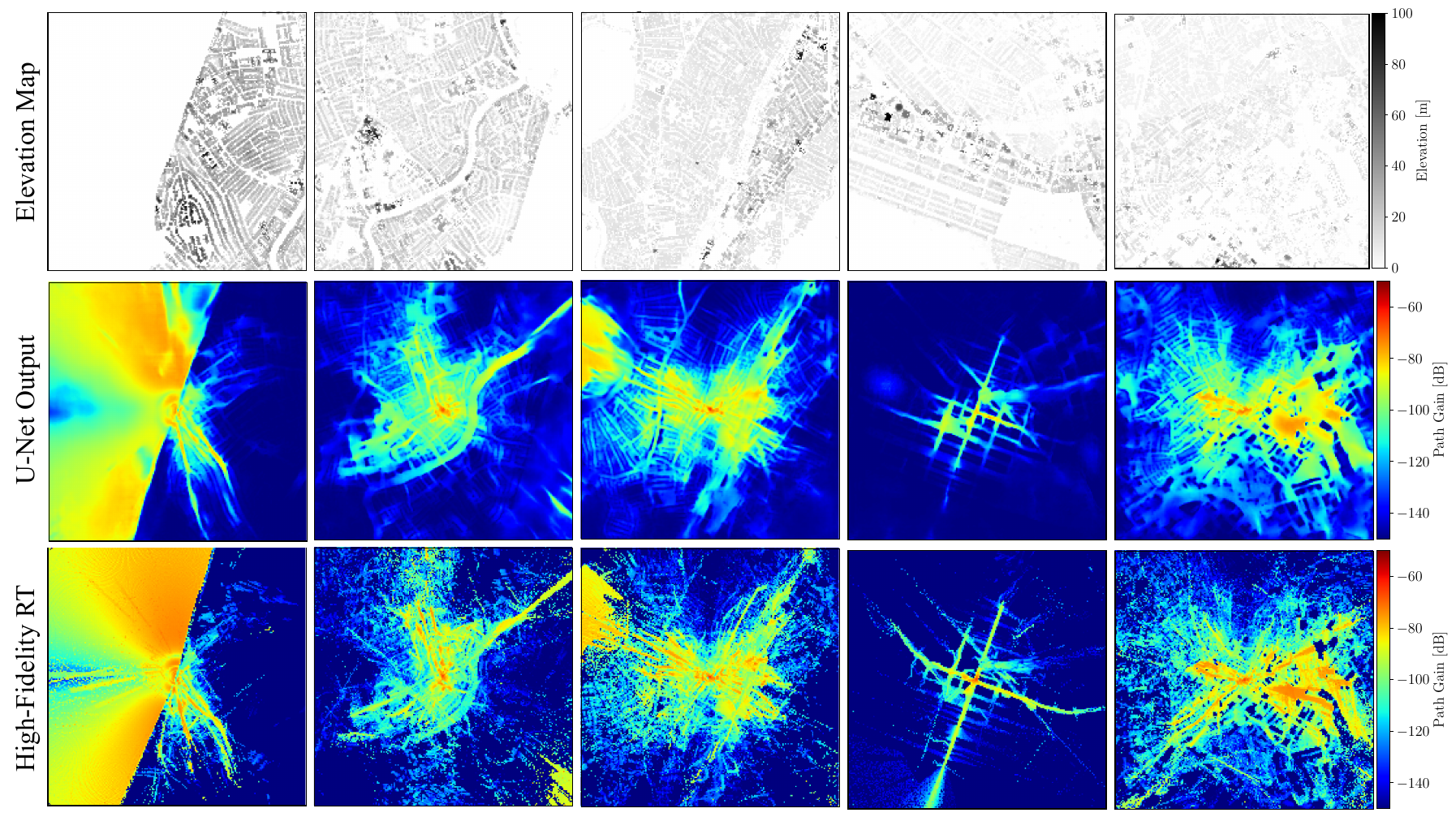}
    \caption{Visual comparison of radio maps across scenarios, ranging from open to dense urban areas. The top row displays the 2D elevation map inputs, the middle row presents the U‑Net model’s predictions, and the bottom row shows the corresponding high‑fidelity ray‑tracing.}
    \label{fig:samples}
\end{figure*}

\review{}
\subsection{U-Net Architecture for Radio Map Estimation}  
\label{sec:unet}
We propose a \gls{dl} framework for estimating radio maps from environmental context, comprising separate training and deployment phases (Fig.~\ref{fig:ai_emulation_process}). During training, the model learns to generate a spatial map of a target channel parameter—such as path gain or RMS delay spread—for any specified \gls{tx} location. Unlike our previous approach~\cite{saeizadeh_ai-assisted_2024}, which required two inputs (an initial rough radio‑map estimate plus an elevation map) and incurred significant overhead, the new model uses only a single‐channel 2D elevation image encoding terrain and building heights. The architecture extends an autoencoder inspired by PMNet~\cite{lee2023scalable}, with adjusted encoder and decoder convolutional layers to process this one‑channel input in place of PMNet’s original dual‑input design.
The network structure is shown in Fig.~\ref{fig:net_arch}, with architectural details provided in Table~\ref{tab:net_arch}.


%
Choosing an effective input space is critical for enabling the model to accurately learn channel characteristics. Prior work has explored a variety of input encoding strategies. For example, Lee et al.\cite{lee2023scalable} used two input images: a building height heat map and a one-hot encoded image indicating the \gls{tx} location. In contrast, Bakirtzis et al.\cite{bakirtzis2022deepray} incorporated richer environmental descriptors such as conductivity, permittivity, relative distance, and free-space path loss maps, primarily for indoor scenarios.

In our approach, we define the input space as $\textbf{x} = \textbf{I}_{el}$, where $\textbf{I}_{el}$ represents a 2D elevation map derived from the scenario’s 3D model. This map encodes terrain elevation, building heights, and other physical obstructions. Although the model input has a fixed spatial dimension (e.g., $200 \times 200$ pixels), we vary the spatial resolution of each sample—ranging from 2.5\,m/pixel to 15\,m/pixel—so that the corresponding physical area spans from 500\,m to 3\,km per side. This resolution-adaptive approach enables the model to learn propagation characteristics over different deployment scales.

The elevation map for our scenario is depicted in Fig.~\ref{fig:samples}. Through experimental analysis, we found that applying min-max normalization, which scales all values between 0 and 1, enhances model convergence. Additionally, we invert the elevation values so that taller buildings—being the primary obstructions in the coverage map—are assigned values closer to zero. This transformation ensures that the model learns the impact of major blockages more effectively.

Ensuring that the model accurately interprets the \gls{tx} location is crucial. Our literature review identified two effective methods for encoding this information: (1) representing the \gls{tx} location as a one-hot-encoded heat map, where a single pixel in the image corresponds to the \gls{tx} position on the building map, and (2) centering the \gls{tx} within the input image, ensuring that the building map is always aligned around the \gls{tx} location. For this work, we adopt the latter, as described in Section~\ref{sec:dataset}.

\subsection{Experimental Results and Analysis}

\begin{figure}[h]
    \centering
    \includegraphics[width=\linewidth]{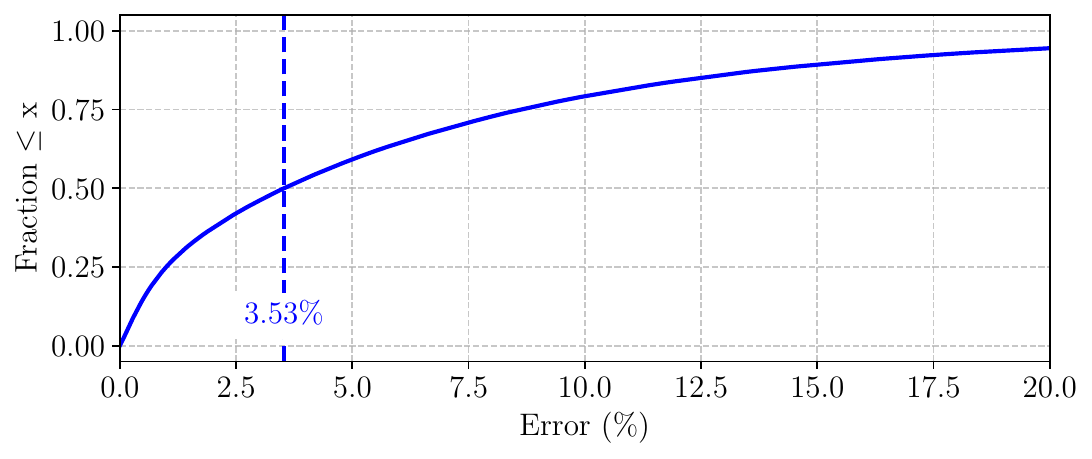}
    \caption{\review{\gls{ecdf} of model performance on the test data, showing a median error of only 3.53~\% across all coverage maps. Radio map values range from -150 dB to -50 dB.}}
    \label{fig:ecdf}
\end{figure}

\review{
For model evaluation, we employ 5-fold cross-validation across the entire dataset. Within each fold, the data is partitioned into training, validation, and testing subsets following a 70-15-15 split ratio, respectively. To prevent data leakage, we ensure that augmented versions of the same scenario are excluded from appearing in multiple subsets. Additionally, the study area shown in Fig.~\ref{fig:boston_map} is divided into a 10$\times$10 grid, with grid cells assigned exclusively to either sets. This spatial partitioning strategy ensures that geographically proximate samples do not appear in both sets, thereby providing a more rigorous evaluation of the model's generalization capability.
}
As shown in Fig.~\ref{fig:ecdf}, the median error percentage across all folds and coverage maps remains around 3.53\%, demonstrating the model's high accuracy. The path gain values in the test dataset range from -150 dB to -50 dB, encompassing a wide variety of propagation conditions. Despite this large dynamic range, the model maintains consistent performance with minimal deviations across different test cases, highlighting its robustness in estimating radio maps with exceptionally low inference error in real-time. Additionally, Fig.~\ref{fig:samples} presents qualitative results showcasing the model's performance across various scenarios, which were randomly selected from the Boston area.

\begin{figure}[t]
    \centering
    \begin{subfigure}[t]{0.98\linewidth}
        \centering
        \includegraphics[width=\linewidth]{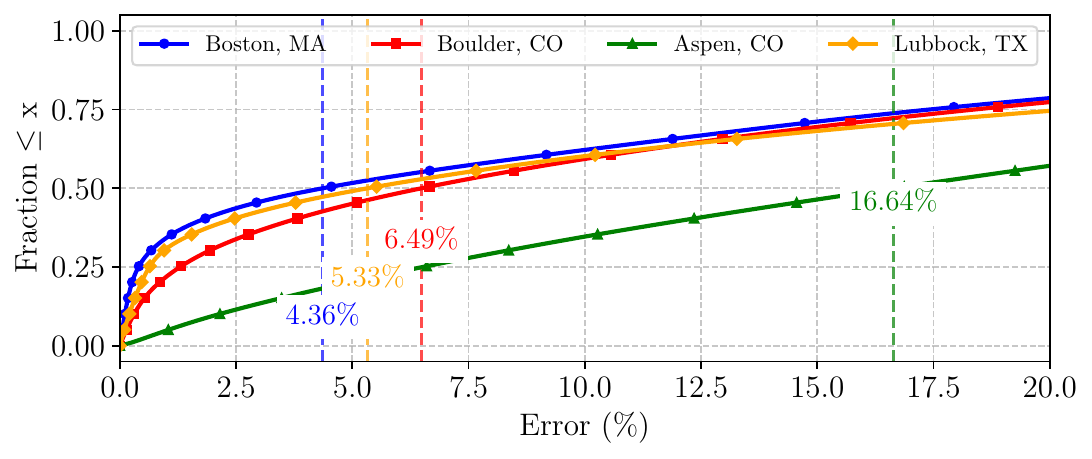}
        \caption{Empirical CDF of prediction error across test regions.}
        \label{fig:ecdf_gen_ecdf}
    \end{subfigure}
    \begin{subfigure}[t]{0.98\linewidth}
        \centering
        \includegraphics[width=\linewidth]{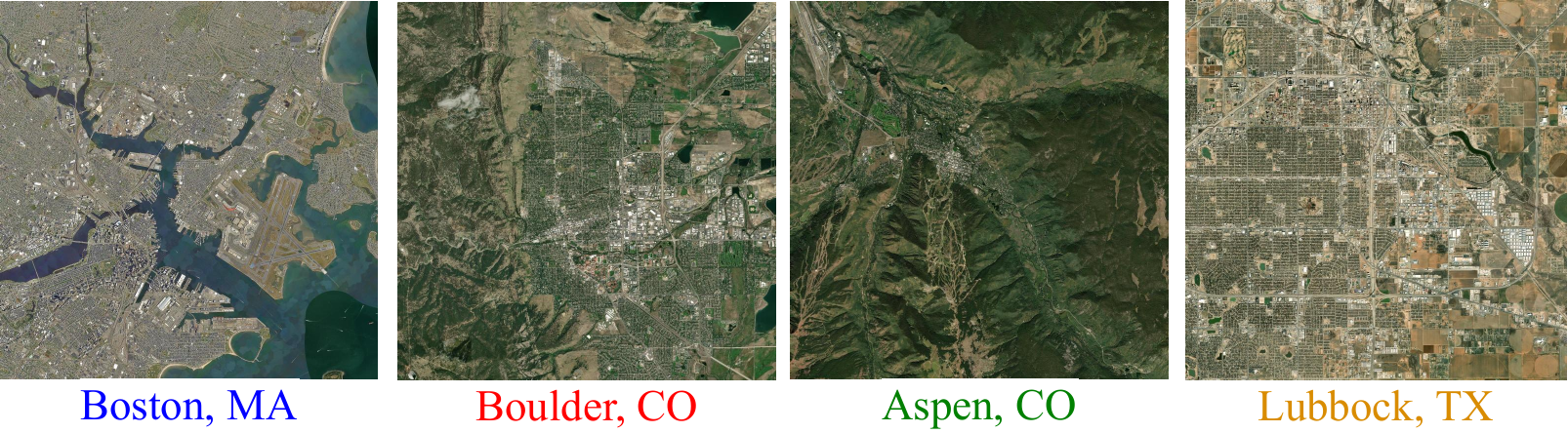}
        \caption{Satellite views illustrating the geographic and morphological diversity of the evaluated regions.}
        \label{fig:ecdf_gen_sat}
    \end{subfigure}
    \caption{\review{Generalization performance of AIRMap across diverse geographic regions. (a) \gls{ecdf} of prediction error for Boston, MA (coastal urban), Boulder, CO (mountainous urban), Aspen, CO (mountainous rural), and Lubbock, TX (flat rural), evaluated using independently sourced environmental data from OpenStreetMap and AWS S3 public elevation tiles.}}
    \label{fig:ecdf_gen}
\end{figure}

\review{
To evaluate the generalization capability of the proposed model, we conducted experiments (only testing without any calibration or retraining) on geographically and morphologically diverse regions using environmental data sourced independently from the training dataset. Specifically, building footprints were obtained from OpenStreetMap, while terrain elevation data were derived from AWS S3 public elevation tiles. Four distinct study areas were selected to represent a range of urban densities and topographical characteristics: (i) \textbf{Flat coastal terrain with high-density urban development:} Boston, MA;  
(ii) \textbf{Dense urban structures in mountainous terrain:} Boulder, CO;  
(iii) \textbf{Rural mountainous region:} Aspen, CO;  
(iv) \textbf{Flat rural terrain:} Lubbock, TX.
 The results, presented in Fig.~\ref{fig:ecdf_gen}, demonstrate that the model maintains sub-10\% median error across most test regions, despite the use of heterogeneous data sources. As anticipated, the Boston scenario exhibits marginally higher error due to discrepancies between the OpenStreetMap-derived geometry and the BostonTwin model used during training. The Aspen scenario yields the highest error, which is attributable to the model's limited exposure to rural mountainous environments during training. These findings confirm that AIRMap generalizes effectively across diverse geographic and morphological conditions, while also identifying domain gaps that warrant further investigation.
}

\review{
The proposed model comprises approximately 37.6 million trainable parameters with a memory footprint of 144.08\,MB, reflecting sufficient capacity to learn complex spatial patterns inherent in radio propagation phenomena. To evaluate computational efficiency, we conducted inference benchmarks over 1,000 forward passes on an NVIDIA L40S GPU. The model achieves an average inference latency of 4.2\,ms per sample, corresponding to a throughput of 238\,samples/s, with a peak GPU memory utilization of 460\,MB and an energy consumption of approximately 1.52\,J per inference. These performance characteristics demonstrate that AIRMap is well-suited for real-time radio map estimation in wireless digital twin applications, satisfying the sub-10\,ms latency requirements for dApp-level operations in O-RAN architectures while maintaining a computationally efficient footprint suitable for large-scale deployment.
}

\subsection{Model Calibration}
\label{sec:model_calibration}

It is important to note that the ray-tracing simulations used for training were conducted at a frequency of 1 GHz (whereas measurement frequency was 933 MHz), employing an isotropic antenna pattern. Additionally, the environmental geometry, while detailed, lacks full accuracy compared to the real-world scenario. Furthermore, the material properties assigned in the simulations differ from actual environmental materials, introducing discrepancies between simulated and measured propagation data. These factors contribute to the simulation-to-reality gap that our model aims to bridge.

To address these discrepancies and improve the real-world accuracy of our \gls{ai}-driven model, we introduce a calibration pipeline for fine-tuning. The model is first pretrained on simulated radio maps to learn general channel characteristics. It is then fine-tuned using a small set of measurements to adapt to real-world conditions while preserving its generalization capabilities.

\review{
To address simulation-to-reality discrepancies and improve real-world accuracy, we introduce a calibration pipeline that fine-tunes the pretrained model using a subset of field measurements. To ensure a realistic assessment of spatial generalization, we adopt a geographically-aware train/test split that clusters measurement points by spatial region, thereby enforcing separation between training and testing areas. This approach is particularly appropriate for geographic data, as it evaluates the model's ability to generalize across space rather than interpolate between nearby points. 
}

To incorporate sparse field measurements into the pretrained model, we employ an analytical calibration approach that computes an optimal affine transformation in closed form. Given the pretrained model's predictions $\hat{P}$ and sparse measurements $P_m$ at $N$ measurement locations, we seek scale $\alpha$ and offset $\beta$ parameters that minimize the squared error:

\review{
\begin{equation}
    \min_{\alpha, \beta} \sum_{i=1}^{N} \left(\alpha \hat{P}_i + \beta - P_{m,i}\right)^2
\end{equation}
The closed-form solution is obtained via ordinary least squares:
\begin{equation}
    \alpha = \frac{\text{Cov}(\hat{P}, P_m)}{\text{Var}(\hat{P})}, \quad \beta = \bar{P}_m - \alpha \bar{\hat{P}}
\end{equation}
}

where $\bar{\hat{P}}$ and $\bar{P}_m$ denote the sample means. The calibrated prediction is then $P_{\text{cal}} = \alpha \hat{P} + \beta$.

\review{
This approach offers several advantages over gradient-based fine-tuning: (i) it requires no iterative optimization, (ii) it is robust to overfitting with extremely sparse data, and (iii) it preserves the spatial patterns learned during pretraining while correcting for systematic bias between simulation and measurement domains.
}

\begin{figure}
    \begin{center}
    \includegraphics[width=0.45\textwidth]{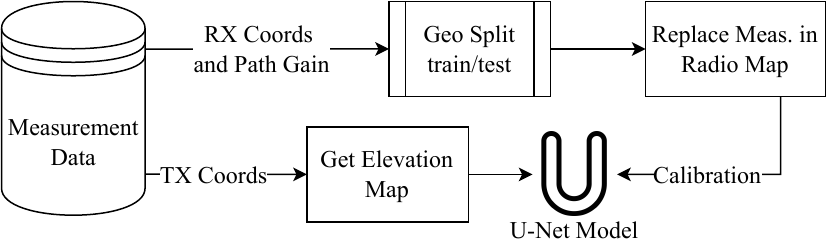}
    \caption{Calibration pipeline to align simulation with measurement data.}
    \label{fig:calibration_process}
    \centering
    \end{center}
\end{figure}

\review{
This strategy enables efficient fine-tuning with minimal data, significantly improving predictive performance. The full calibration process is illustrated in Fig.~\ref{fig:calibration_process}.
}

\begin{figure}
    \centering
    \includegraphics[width=0.48\textwidth]{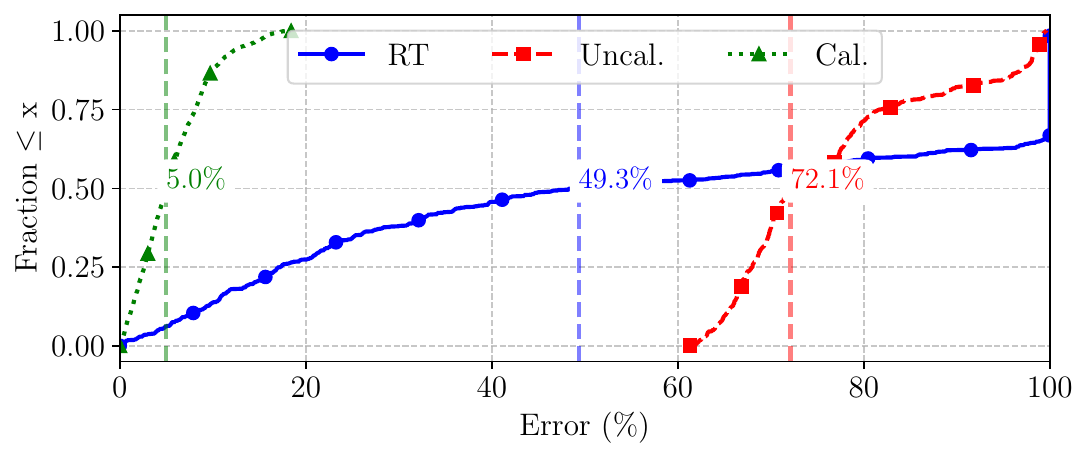}
    \caption{\review{\gls{ecdf} of error with and without calibration, using only 20\% of the measurement dataset for training.}}
    \label{fig:ecdf_calibration}
\end{figure}

\review{
We perform the calibration process 100 times to account for randomness in the train-test splits, ensuring that the results are not biased toward any specific subset of the data. Fig.~\ref{fig:ecdf_calibration} presents a performance comparison among Sionna RT, the calibrated U-Net model, and the uncalibrated model. As shown, the calibrated model consistently outperforms the others, reducing the median error to approximately 5\% and producing predictions that more closely align with real-world measurements.
}

\begin{figure}
    \centering
    \includegraphics[width=0.48\textwidth]{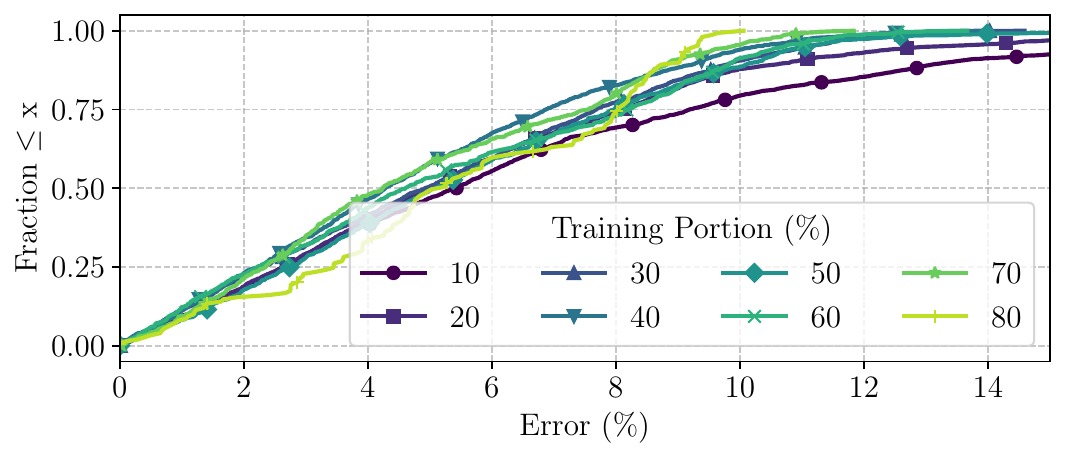}
    \caption{\review{\gls{ecdf} of the error for different proportions of measurement data used to train the calibration process.}}
    \label{fig:ecdf_split_ratio}
\end{figure}

\review{
The effect of varying the calibration data ratio is illustrated in Fig.~\ref{fig:ecdf_split_ratio}. As shown, increasing the proportion of measurement data used for calibration yields only marginal improvements in accuracy. For instance, using 20\% of the measurements achieves a median error of approximately 5\%, whereas using 80\% reduces the error to 3.57\%---an improvement of only 1.43 percentage points. This trade-off must be considered in the context of data collection costs and availability. Furthermore, as discussed in Section~\ref{sec:digital_twin}, errors of this magnitude do not significantly impact system-level performance, suggesting that 20\% represents a practical and cost-effective calibration threshold.
}
 
This calibration pipeline allows the U‑Net model to continuously incorporate real‑world measurements, ensuring that the digital twin remains closely aligned with the physical network. By iteratively collecting field data, updating the model, and regenerating the radio maps, the system maintains high fidelity and accuracy for subsequent ``what‑if'' analyses and control actions.

Despite Sionna RT’s support for gradient‑based calibration of material and antenna parameters~\cite{hoydis2023learning}, we did not apply this process to our comparisons for two main reasons. First, the calibration methods in Sionna RT rely on full \gls{cir} and differentiable parametrizations of scattering and antenna patterns, whereas our measurements supply only scalar path‑gain values. Without access to absolute delays, phases, or multi‑tap \gls{cir} data, Sionna’s calibration pipeline cannot be directly applied. Second, the end‑to‑end differentiable calibration in Sionna RT incurs substantial computational overhead—requiring repeated ray tracing and backpropagation through complex computational graphs—which contradicts our real‑time objectives. In contrast, our lightweight U‑Net calibration uses only sparse path‑gain samples and executes in milliseconds, ensuring both agility and fidelity for digital‑twin applications.

\section{Radio Maps for Real-Time Digital Twins}
\label{sec:digital_twin}

\begin{figure}
    \centering
    \includegraphics[width=0.45\textwidth]{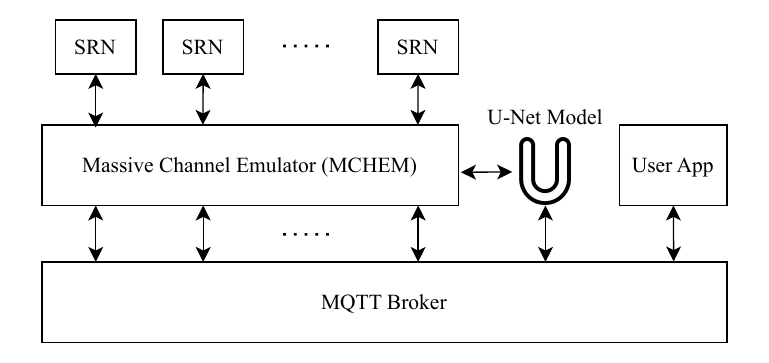}
    \caption{Real-time channel emulation pipeline on Colosseum. User-selected locations trigger U-Net inference, with predicted taps sent via MQTT to \gls{mchem} for emulation.}
    \label{fig:mqtt}
\end{figure}

\begin{figure}[!ht]
    \centering
    \vspace{-2pt}
    \begin{subfigure}[b]{0.49\textwidth}
        \centering
        \includegraphics[width=\textwidth]{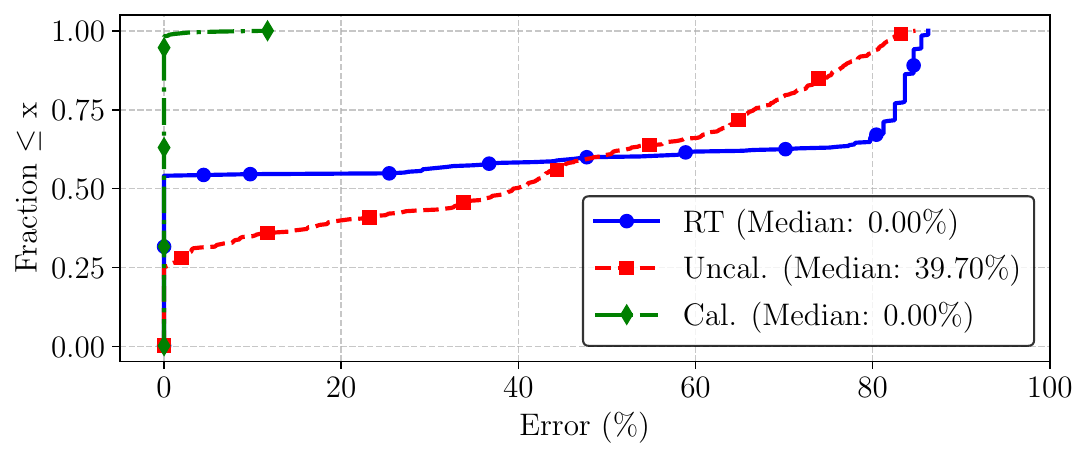}
        \caption{ECDF of Shannon capacity error}
        \label{fig:shannon}
    \end{subfigure}
    \begin{subfigure}[b]{0.49\textwidth}
        \centering
        \includegraphics[width=\textwidth]{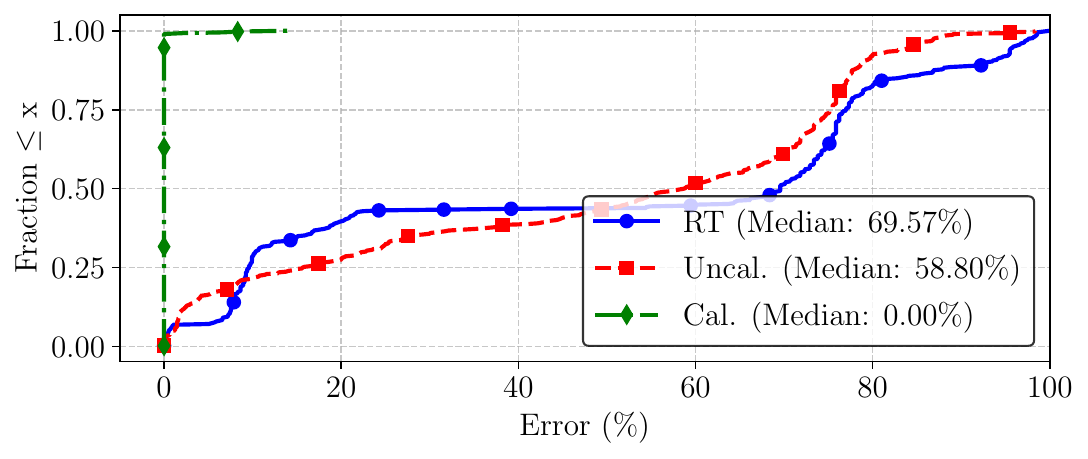}
        \caption{ECDF of link‐adaptation efficiency error}
        \label{fig:la_spectral}
    \end{subfigure}
    \begin{subfigure}[b]{0.49\textwidth}
        \centering
        \includegraphics[width=\textwidth]{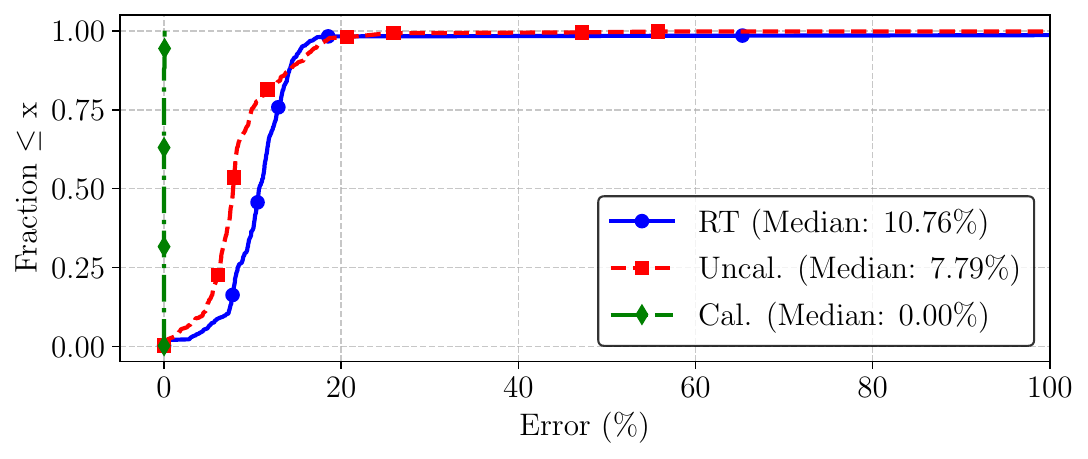}
        \caption{\gls{ecdf} of \gls{bler} error}
        \label{fig:bler}
    \end{subfigure}
    \caption{Error distribution ECDFs comparing measurement data, Sionna RT, and both uncalibrated and calibrated U-Net models across (a) Shannon capacity, (b) link-adaptation efficiency, and (c) block‐error rate.}
    \label{fig:system_level_comparison}
\end{figure}

\review{
As introduced in Section~\ref{sec:intro}, the goal of these radio maps is to enable real-time estimation of channel parameters over a geographic area for wireless digital twins—real-time, scalable simulators that replicate the behavior of actual wireless networks. The next step is to integrate our U-Net model into higher-level workflows for system-level evaluation. In this section, we demonstrate and evaluate how the U-Net functions as a channel-parameter generator for both Sionna SYS system simulations~\cite{sionna} and Colosseum—the world's largest wireless network emulator~\cite{polese2024colosseum}.
To ensure a consistent and fair comparison, all system-level experiments were conducted using the same trajectory and measurement described in Section~\ref{sec:model_reality}, enabling direct evaluation of AIRMap's performance against empirical ground-truth data.
}

\subsection{Evaluation on Sionna SYS}

Sionna SYS provides a modular framework to simulate full-stack system-level behavior. In each time slot, the system performs user scheduling across the resource grid, allocates transmit power to each user, computes the \gls{sinr}, selects the \gls{mcs} via link adaptation, and generates decoded bits along with \gls{harq} feedback using physical layer abstraction.

In this framework, we assume perfect channel state information for precoder and equalizer design, achievable‐rate estimation (for scheduling decisions), and channel‐quality feedback (for link adaptation). We evaluate performance using both Shannon capacity and \gls{olla}‑adjusted spectral efficiency, where \gls{olla} dynamically selects the optimal \gls{mcs} level.

We evaluate how path gain prediction errors impact system-level performance, focusing on spectral efficiency and \gls{bler}, as well as the feasibility of real-time simulation. For this analysis, we use the Sionna SYS simulator~\cite{sionna}, which enables realistic wireless network simulations through a flexible and Python-based interface.

Fig.~\ref{fig:system_level_comparison} compares errors for three channel‑modeling approaches—Sionna RT, the uncalibrated U‑Net, and the calibrated U‑Net—against measurement data. Each model employs a single‑tap channel with fixed propagation delay and randomized phase to focus exclusively on path‑gain effects in system‑level evaluations.

The results show that, despite a 10\% error in path‑gain estimates, the calibrated U‑Net achieves near-zero error across system‑level metrics when compared to measurement‑based channels. The uncalibrated U‑Net also outperforms the ray‑tracing baseline. These outcomes confirm that the U‑Net’s residual path‑gain deviations do not impact higher‑layer simulations, validating its applicability for real‑time system‑level wireless emulation.

We implemented the full pipeline—from elevation map to system-level metrics—using Sionna SYS on an NVIDIA L40S GPU. After warm-up, each step is processed in 0.75\,ms, and full radio map generation (200×200 points) takes 280\,ms (284\,ms including radio map processing), enabling real-time operation. The pipeline leverages TensorFlow XLA with mixed precision and batch parallelism, achieving 14,005 steps/sec/point for rapid adaptation in dynamic wireless environments.

\begin{figure*}
    \centering
    \includegraphics[width=0.95\linewidth]{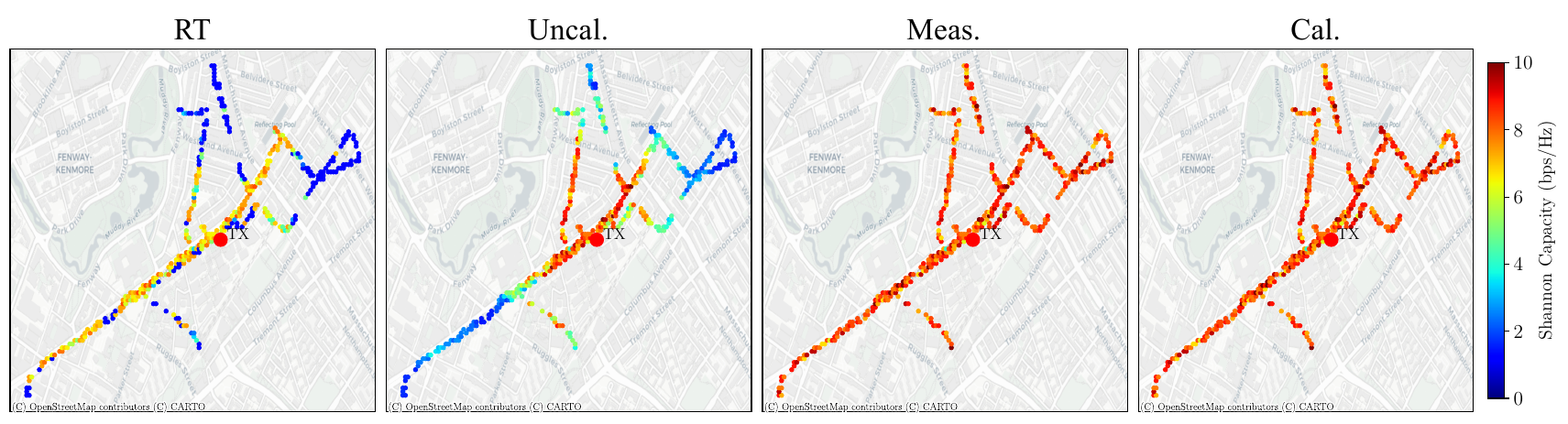}
    \caption{Shannon capacity heatmaps for ray tracing, uncalibrated U‑Net, measurement data, and calibrated U-Net model estimations on test measurement points.}
    \label{fig:shannon_map}
\end{figure*}

\subsection{Deployment on Colosseum}

To enable real-time channel emulation as well as simulation using our U-Net model, we integrated it into the Colosseum platform. We followed the approach introduced in ColosSUMO~\cite{gemmi_colossumo_2024}, which extends Colosseum with real-time scenario generation capabilities. As shown in Fig.~\ref{fig:mqtt}, when a user selects or updates a location, the corresponding building elevation map is loaded, transformed into the appropriate input shape and format within 20\,ms, and passed to our U-Net model. The model performs inference in approximately 4\,ms, producing a path gain prediction. This result is then sent to an MQTT broker, which forwards it to Colosseum's \gls{mchem}, the FPGA-based Massive Channel Emulator. \gls{mchem} then uses these taps to emulate the corresponding wireless channel in real time.

\begin{figure}
    \centering
    \includegraphics[width=0.48\textwidth]{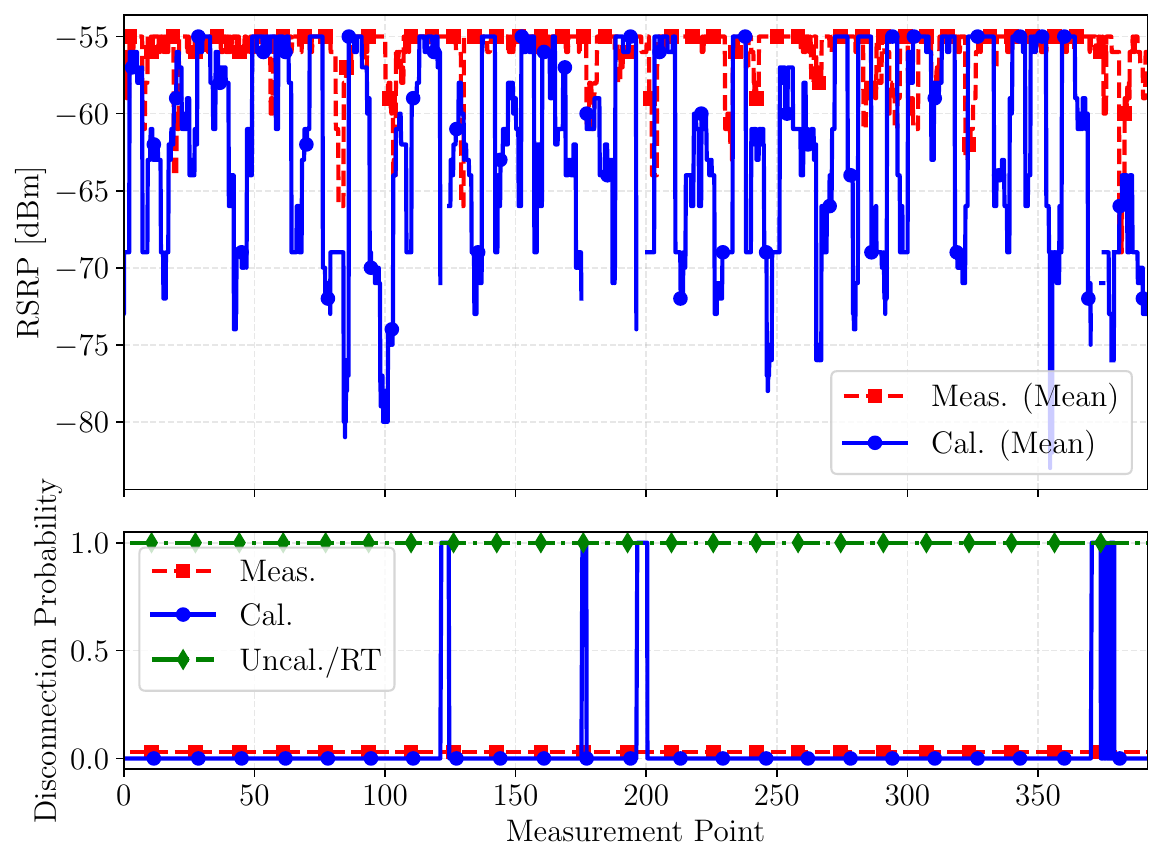}
    \caption{Comparison of \gls{rsrp} reports from \gls{oai} on Colosseum (10 runs per experiment). The Calibrated U-Net closely matches measurements, while Ray Tracing and Uncalibrated U-Net failed to establish a connection in this setup.}
    \label{fig:rsrp_oai}
\end{figure}

Fig.~\ref{fig:rsrp_oai} reports the measured RSRP from running \gls{oai} on the Colosseum testbed using single-tap time-domain channels with simple propagation delays to simulate mobile scenarios across four channel models: Ray Tracing, Calibrated U-Net, Uncalibrated U-Net, and actual measurements. Each experiment was repeated 10 times. Both the Ray Tracing and Uncalibrated U-Net models failed to establish a connection under the same configuration, resulting in no \gls{rsrp} data. In contrast, the Calibrated U-Net model successfully produced \gls{rsrp} patterns closely matching empirical measurements, with an \gls{rmse} of 8.86\,dBm.
%

\section{Conclusions}
\label{sec:conclusion}

We have presented AIRMap, an AI-driven framework for real-time radio-map estimation that bridges the gap between high-fidelity ray tracing and fast, scalable inference. By relying solely on a novel single-channel elevation map—which is simple, cost-effective, and widely accessible—as input, and a U-Net–based autoencoder, AIRMap produces accurate path-gain maps in under 4\,ms. Its data-driven design allows efficient calibration using minimal field measurements, enabling site-specific adaptation and correction of residual errors. Extensive evaluation on a large Boston-area dataset shows AIRMap matches measurement-based performance on key system-level metrics, such as spectral efficiency and \gls{bler}. Integration into the Colosseum emulator and the Sionna SYS platform confirms AIRMap's practical feasibility for real-time, end-to-end wireless network emulation.

A key feature of AIRMap is its resolution-adaptive input representation: while the model accepts fixed-size elevation map inputs, it supports variable spatial resolutions, enabling radio-map estimation across physical areas ranging from 500\,m to 3\,km per side without altering the architecture. This design supports flexible deployment scenarios—from dense urban zones to wider suburban or campus-scale regions—while preserving inference speed and model accuracy.

AIRMap's contributions lie in drastically reducing computational cost and latency for radio environment modeling while using only easily obtainable elevation maps. Its data-driven nature allows for seamless incorporation of realistic wireless channel effects into real-time network management, adaptive resource allocation, and rapid scenario testing. These capabilities empower researchers and network operators to dynamically optimize system performance and accelerate the development of advanced digital twin applications requiring fast, accurate propagation predictions from minimal input data.


\bibliographystyle{IEEEtran}
\bibliography{biblio.bib}

\vspace{-20pt}

\begin{IEEEbiography}[{\includegraphics[width=1in,height=1.25in,clip,keepaspectratio]{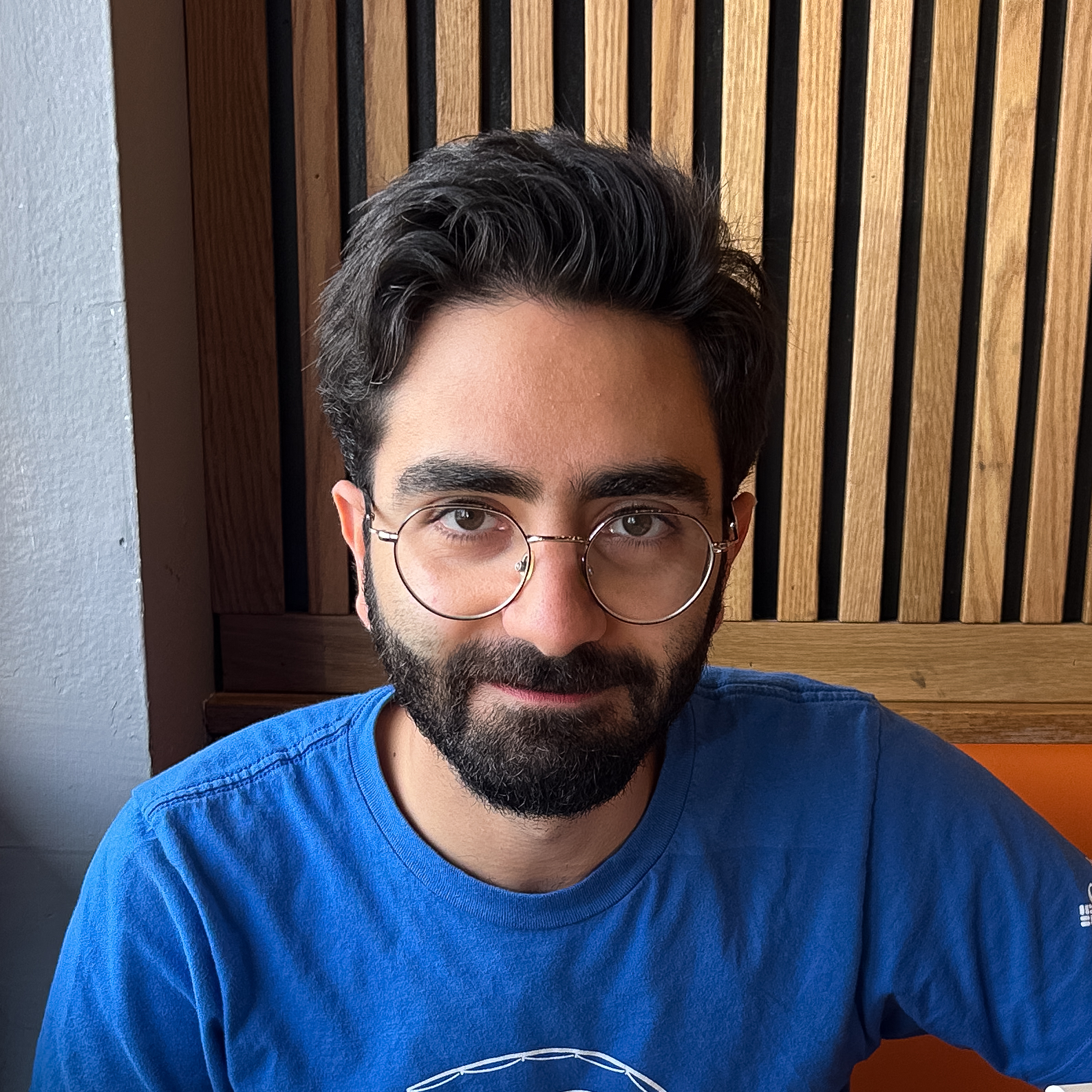}}]{Ali Saeizadeh}
is a Ph.D. candidate at the Institute for Intelligent Networked Systems at Northeastern University, USA, advised by Prof. Tommaso Melodia. He received his B.S in Electrical Engineering from the University of Tehran in 2022, and M.S. from Northeastern University, Boston, MA in 2026. His research focuses on applying AI/ML to wireless communication systems, Wireless Digital Twins, 5G/6G networks, channel modeling. 
\end{IEEEbiography}

\vspace{-20pt}

\begin{IEEEbiography}[{\includegraphics[width=1in,height=1.25in,clip,keepaspectratio]{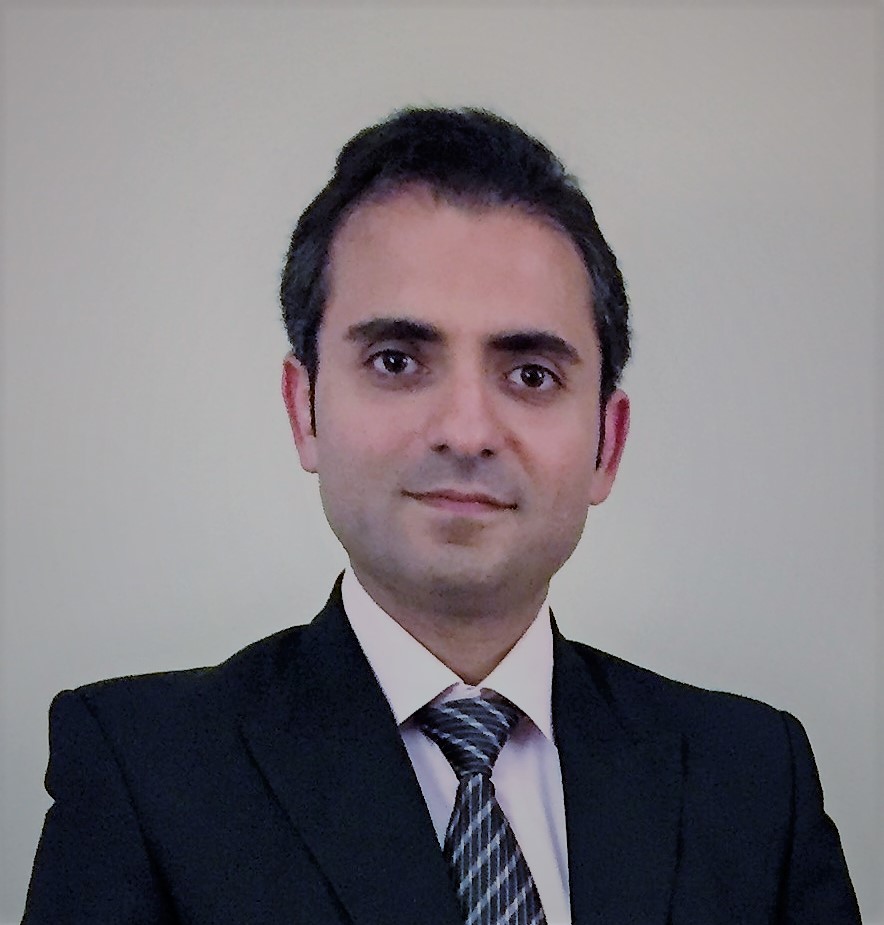}}]{Miead Tehrani-Moayyed}
is a Ph.D. candidate in Computer Engineering at Northeastern University in Boston. He is working under Prof. Stefano Basagni's supervision on RF channel models for static and mobile scenarios, from simulations to models for large-scale emulations. His research interests include applying AI/ML algorithms to wireless communication, propagation models for next-generation cellular systems, and computer networks. He received his M.S. in Computer Systems Architecture Engineering from Azad University, Iran in 2013 and his B.S. in Computer Engineering from Shomal University, Iran in 2007.
\end{IEEEbiography}

\vspace{-20pt}

\begin{IEEEbiography}[{\includegraphics[width=1in,height=1.25in,clip,keepaspectratio]{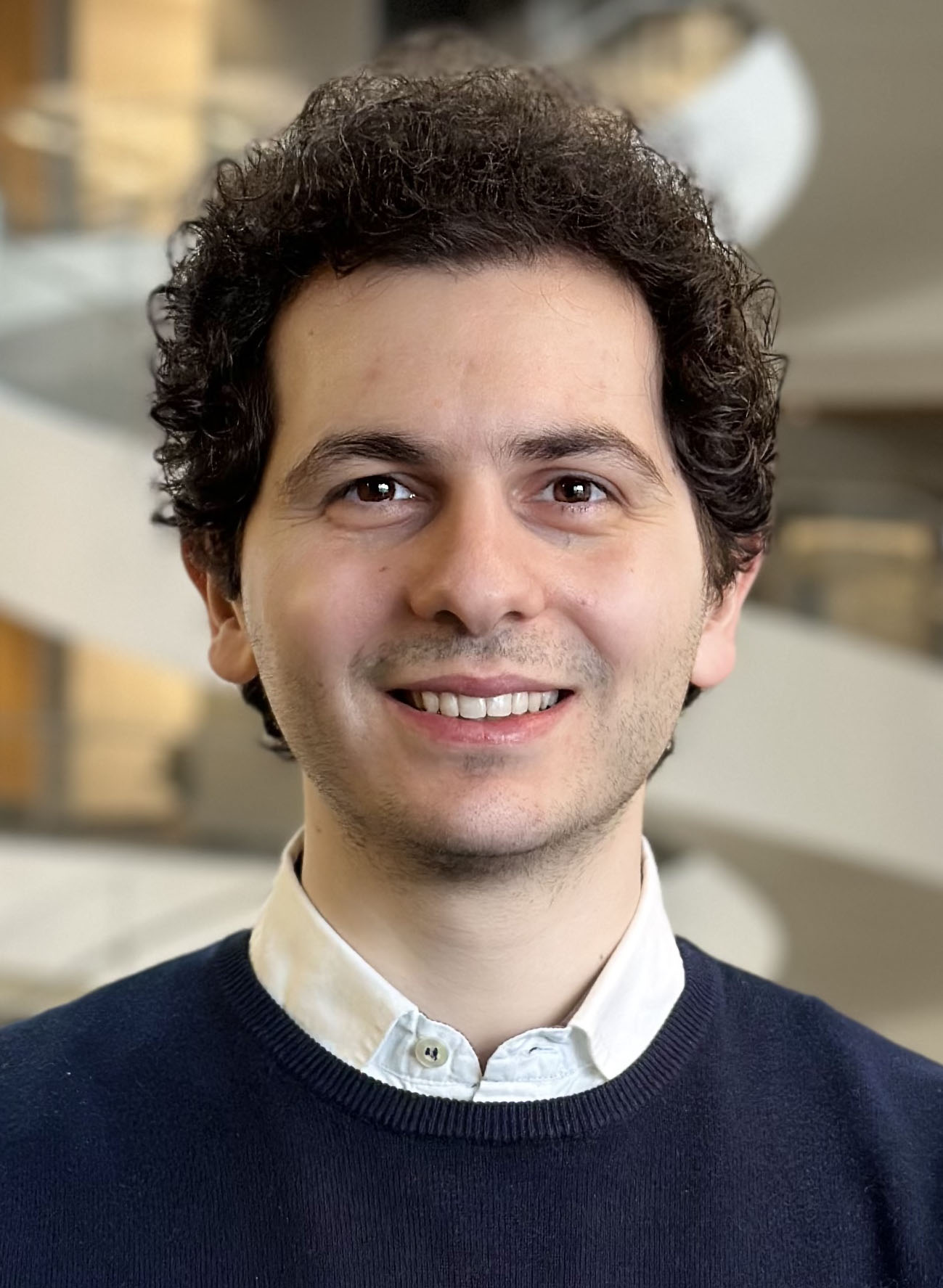}}]{Davide Villa} is a senior GPU software engineer at NVIDIA. He received his B.S. and M.S. in Computer Engineering from the University of Pisa, Italy, in 2015 and 2018, respectively, and his Ph.D. in Computer Engineering from Northeastern University in 2025. From 2018 to 2020, he was a Research Scientist in the Embedded Systems and Network Group at United Technologies Research Center in Cork, Ireland. His research interests include 5G-and-beyond cellular networks, channel characterization for wireless systems, O-RAN, and software-defined networking for wireless networks.
\end{IEEEbiography}

\vspace{-20pt}

\begin{IEEEbiography}[{\includegraphics[width=1in,height=1.25in,clip,keepaspectratio]{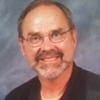}}]{J. Gordon Beattie, Jr.} is a 
Senior Principal Research Scientist/Engineer at VIAVI Solutions. His focus in recent years has been in the areas of wireless communications systems and IP network optimization, RF interference identification and mitigation, and Electromagnetic Compatibility management. Much of his patentable work has focused in these areas where he is a recognized corporate leader in the optimization of 6G, 5G, 4G/LTE, Land Mobile Radio and ADSL/VDSL networks in support of mobile and fixed video streaming, IoT, Aumented/VirtualReality and their Edge and Cloud Computing architectures.
\end{IEEEbiography}

\vspace{-20pt}

\begin{IEEEbiography}[{\includegraphics[width=1in,height=1.25in,clip,keepaspectratio]{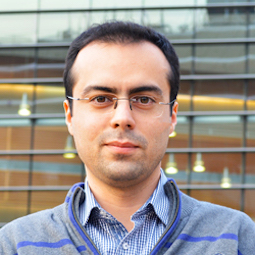}}]{Pedram Johari}
is a Principal Research Scientist at Northeastern University. His background and research interests are in the fusion of AI and future generation of cellular networks and Open RAN, in particular focused on enabling full-protocol real-time digital twins for research and development, system integration and testing for open, programmable and AI-driven wireless networks.
He serves as the lead and co-Principal Investigator on multiple grants from U.S. federal agencies including the NSF, USDOT, OUSD(R\&E), and NTIA Public Wireless Supply Chain Innovation Funds, as well as multiple industry sponsored projects. Pedram received his Ph.D. in Electrical Engineering from the University at Buffalo, NY, 2018, and his MBA from the D’Amore-McKim School of Business at Northeastern University, Boston MA, 2024.
Pedram 
is a member of the IEEE and ACM, and has collaborated with several academic and industrial research partners in multiple Open RAN related projects. He is the Editor-in-Chief of the Elsevier Software Impacts journal.
\end{IEEEbiography}

\vspace{-20pt}

\begin{IEEEbiography}
[{\includegraphics[width=1in,height=1.25in,keepaspectratio]{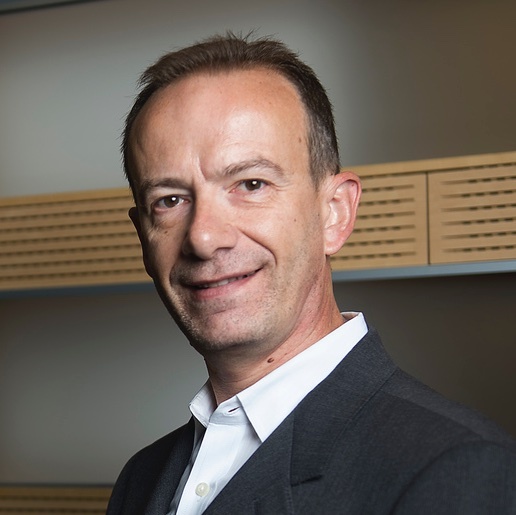}}]{Stefano Basagni}
is with the Intelligent Networked Systems Institute (INSI) and a professor at the ECE Department at Northeastern University, in Boston, MA. He holds a Ph.D.\ in electrical engineering from the University of Texas at Dallas (2001) and a Ph.D.\ in computer science from the University of Milano, Italy (1998). Dr. Basagni's current interests concern research and implementation aspects of mobile networks and wireless communications systems, wireless sensor networking for IoT (underwater, aerial and terrestrial), and definition and performance evaluation of network protocols.
Dr. Basagni has published over thirteen dozen of highly cited, refereed technical papers and book chapters. His h-index is currently 57 (June 2026). He is also co-editor of three books. Dr. Basagni served as a guest editor of multiple international ACM/IEEE, Wiley and Elsevier journals. He has been the TPC co-chair of international conferences. He is a distinguished scientist of the ACM, a senior member of the IEEE, and a member of CUR (Council for Undergraduate Education).
\end{IEEEbiography}

\vspace{-20pt}

\begin{IEEEbiography}
[{\includegraphics[width=1in,height=1.25in,keepaspectratio]{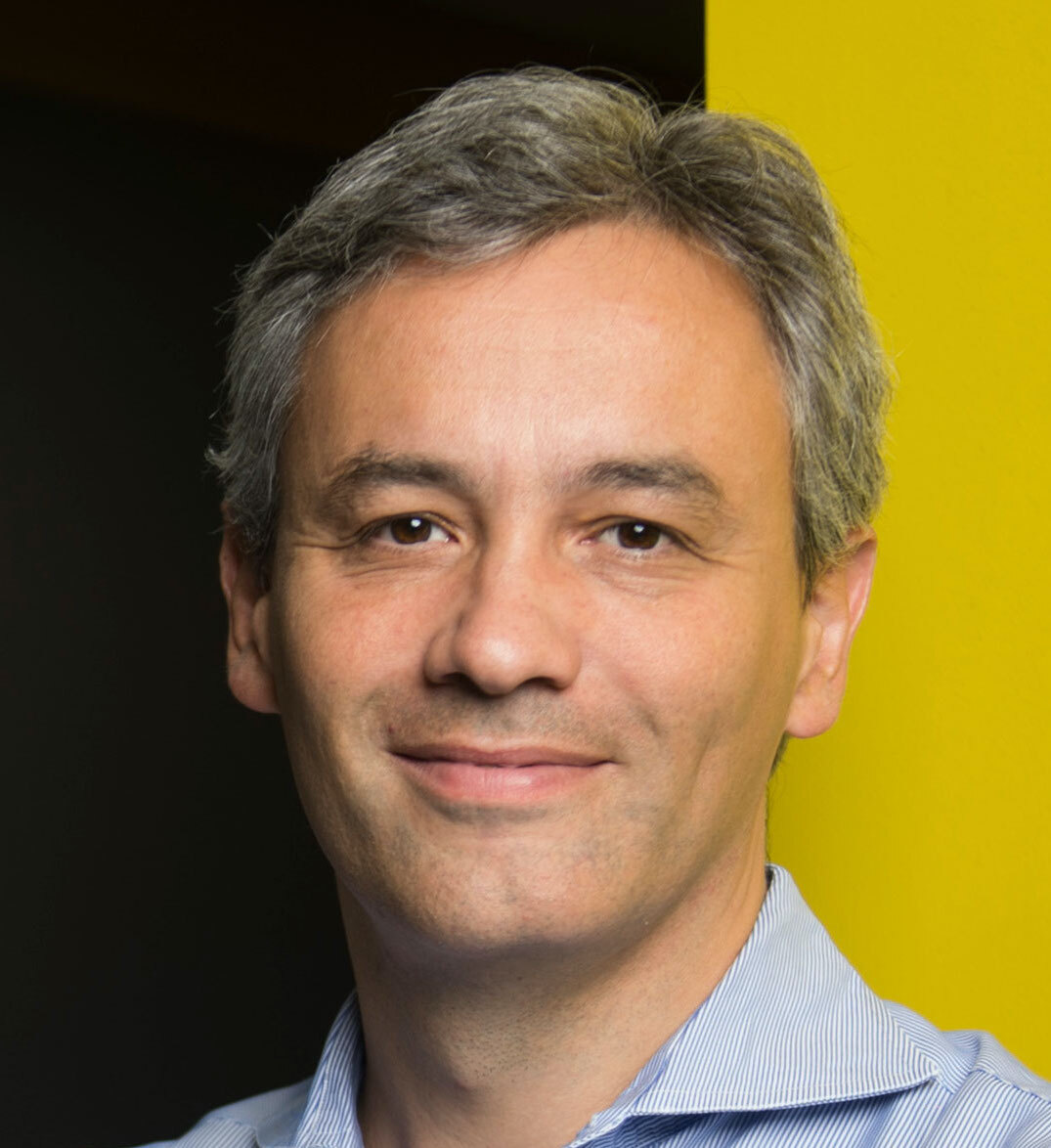}}]{Tommaso Melodia}
is the William Lincoln Smith Chair Professor with the Department of Electrical and Computer Engineering at Northeastern University in Boston. He is also the Founding Director of the Institute for Intelligent Networked Systems and was the Director of Research for the PAWR Project Office. He received his Ph.D. in Electrical and Computer Engineering from the Georgia Institute of Technology in 2007. He is a recipient of the National Science Foundation CAREER award.  Prof. Melodia is the Principal Investigator of Open6G, a cooperative research and development center shaping the future of open, programmable, and AI-powered wireless systems. He sits on the board of the AI-RAN Alliance and of the OpenAirInterface Software Community. Prof. Melodia's research on modeling, optimization, and experimental evaluation of Internet-of-Things and wireless networked systems has been funded by the National Science Foundation, the Air Force Research Laboratory the Office of Naval Research, DARPA, and the Army Research Laboratory. Prof. Melodia is a Fellow of the IEEE, of the ACM, and of the National Academy of Inventors.
\end{IEEEbiography}

\end{document}